\documentclass[aps,prl,reprint,letterpaper,amsmath,amssymb,longbibliography,superscriptaddress]{revtex4-2}
\usepackage{amsmath}
\usepackage{graphicx}

\begin{document}
\title{Selective measurement of the longitudinal electron spin relaxation time $T_1$ of Ce$^{3+}$ ions in a YAG lattice: Resonant spin inertia}
\author{V.~V.~Belykh}
\email[]{belykh@lebedev.ru}
\affiliation{P.N. Lebedev Physical Institute of the Russian Academy of Sciences, 119991 Moscow, Russia}
\author{S.~R.~Melyakov}
\affiliation{P.N. Lebedev Physical Institute of the Russian Academy of Sciences, 119991 Moscow, Russia}

\begin{abstract}
Electron spin oriented along an external magnetic field is subject to longitudinal spin relaxation with characteristic time $T_1$. The corresponding decay is nonoscillating, so one cannot readily ascribe $T_1$ to a certain $g$ factor. This becomes a problem when several electronic states with different $g$ factors are present in the system, e.g. electrons and holes. We solve this problem by optically pumping spin polarization and then selectively depolarizing it using a radio frequency (rf) field. By modulating the rf field amplitude one can observe the retarded modulation of spin polarization which depends on the relation between the modulation period and $T_1$. Using this resonant spin inertia method, we unveil the strong anisotropy of $T_1$ for rare-earth Ce$^{3+}$ ions in a YAG crystal at low temperatures and low magnetic fields. We also show that the spread of Larmor frequencies within the electron ensemble in this system is not static, but results from the fluctuations of internal magnetic fields on a timescale much shorter than $T_1$. 

\end{abstract}
\maketitle

\section{Introduction}
Electron spin dynamics in a magnetic field $\mathbf{B}$ depends on the initial spin state or spin orientation, on the classical language. Spin oriented perpendicular to the magnetic field precesses around $\mathbf{B}$ with the Larmor frequency $\omega_\text{L}$. This transverse precession is also subject to relaxation with time $T_2$, which is also called spin coherence time. The transverse relaxation can either change or not change the spin energy. There are several methods allowing to determine $T_2$ based on spin echo \cite{Hahn1950} and more recent ones based on stimulated resonant spin amplification \cite{BelykhPRL2021} and spin mode locking effects \cite{Greilich2006}. Determining of $T_2$ is always accompanied by dealing with spin precession and measurement its frequency $\omega_\text{L}$, i.e., $g$ factor, which is a fingerprint of the electronic state being addressed. So one knows exactly the correspondence between $g$ factor and $T_2$. 

On the other hand, spin aligned along $\mathbf{B}$ changes its orientation only as a result of relaxation caused by perturbations. This longitudinal spin relaxation is characterized by time $T_1$ and is accompanied by a change of the spin energy. There exist several methods of studying the longitudinal spin relaxation and measuring $T_1$, which are based on the electron spin resonance (ESR) technique \cite{Schweiger2001}. Among them are saturation-recovery \cite{Alger1968}, inversion-recovery \cite{Poole1996} and microwave amplitude modulation \cite{Herve1960,Misra2006} techniques. These methods, being very powerful, suffer from limited sensitivity and can mostly address macroscopic bulk systems. Furthermore, they are mainly applied in high magnetic fields, since the ESR signal is proportional to the equilibrium spin polarization, i.e., to $B$. Meanwhile, some interesting aspects of spin dynamics, e.g., electron-nuclear spin coupling, become especially pronounced in low fields.  There are also optical methods of studying longitudinal spin dynamics: time-resolved photoluminescence \cite{Colton2004,Akimov2006,Fu2006,Colton2007,Linpeng2016,Siyushev2014}, pump-probe Faraday/Kerr rotation \cite{Baumberg1994,Zheludev1994,Belykh2016Ext}, and spin inertia \cite{Heisterkamp2015,Zhukov2018,Smirnov2018}. These methods offer better sensitivity, can be used at low fields and can be applied to various systems provided the electron state of interest can be addressed by photons with energies lying in the range accessible to the experimentalist. However, as far as dynamics of the longitudinal spin component is nonoscillating, but shows monotonic decay with time $T_1$, it gives no information about $g$ factor. To determine $g$ factor one can also measure the precession of the transverse spin component in magnetic field inclined with respect to the optically oriented spin polarization. However, when the system under study contains several spin states, e.g., electron and hole ones \cite{Belykh2019,Kirstein2021}, $T_1$ measured from monotonic decay cannot be readily assigned to a certain $g$ factor from several measured ones. In this work we solve this problem by taking advantage of the sensitivity of optical methods and selectivity of ESR-based methods. 

We pump spin polarization by the laser beam which simultaneously probes spin polarization via Faraday rotation. Also we apply radiofrequency (rf) field which selectively reduces spin polarization corresponding to a given $g$ factor. By modulating the rf field amplitude, we observe the spin inertia effect, i.e., the retarded modulation of the spin polarization. The amplitude of the spin modulation and its retardation depends on the relation between the modulation period and $T_1$. This allows us to determine $T_1$. In contrast to ESR, here the spin polarization is determined by the optical pumping and is much larger than the equilibrium spin polarization. This makes it possible to perform measurements at low magnetic fields where the equilibrium spin polarization is very low.
We apply this method to study the longitudinal spin dynamics of the unpaired electron in the 4f level of rare-earth Ce$^{3+}$ ions in the yttrium aluminum garnet (YAG) lattice. This electron state has a strongly anisotropic $g$ tensor. Since Ce$^{3+}$ ions occupy six magnetically inequivalent sites in YAG lattice, in the experiment one observes six spin resonances corresponding to  different orientations of the $g$ tensor \cite{Lewis1966}. Studying longitudinal spin dynamics by all-optical methods give some average $T_1$, which would make sense for isotropic $T_1$. Here, we are able to measure $T_1$ for each spin resonance individually and reveal strong anisotropy of $T_1$, which changes by more than a factor of 2 while changing the magnetic field orientation. Time $T_1$ has also strong magnetic field dependence at low $B$. Next, we address the nature of the spread in the Larmor frequencies, which determines the width of the ESR spectra. It turned out that the behavior of the spin ensemble subject to a rf field is drastically different depending on whether this spread is caused by frozen (on the timescale of $T_1$) or time-varying fluctuations. We show that Larmor frequencies in the investigated Ce$^{3+}$:YAG system fluctuate on the timescale much shorter than $T_1$.

\section{Experimental details}
The sample under study is a 0.5-mm-thick Ce$^{3+}$:YAG crystal with a Ce$^{3+}$
ion concentration of 0.5 at. \%. The same sample was used in works \cite{BelykhPRL2021,Azamat2017}. The scheme of the experiment is shown in Fig.~\ref{fig:BDep}(a). The sample is placed in a variable temperature ($5-300$~K) He-flow cryostat.  Using a permanent magnet placed outside the cryostat at a controllable distance from the sample, a constant magnetic field $\mathbf{B}$ up to 20~mT is applied; it was measured using a Hall sensor placed in the vicinity of the sample. The magnet is placed on a goniometer to control the magnetic field direction. The optical spin pumping and probing are performed by a same laser beam with elliptical polarization. The circular and linear polarization components of the beam serve as simultaneous pump and probe, respectively, for an electron spin polarization \cite{Belykh2020,BelykhPRL2021}. Experiments with pumping and probing of a spin system using a single beam were also performed previously for atoms \cite{Bell1961} and semiconductors \cite{Saeed2018}. We use a pulsed Ti:Sapphire laser operating at a wavelength of 888~nm that is frequency doubled with a BBO crystal to obtain a wavelength of 444~nm. The laser generates a train of 100-fs-long optical pulses with a repetition frequency of 77~MHz. We note that for the method introduced in this work the fact that the laser be pulsed is not essential; a pulsed laser is used to perform efficient frequency doubling to reach the desired wavelength. The laser beam is focused on a sample onto a 100~$\mu$m spot, which together with the small sample thickness ensures a negligible magnetic field variation within the investigated region of the sample. We measure the spin polarization via the Faraday rotation of the linear polarization component of the laser beam passed through the sample. It is analyzed using a Wollaston prism, splitting the beam into two orthogonally polarized beams of approximately equal intensities that are further registered by a balanced photodetector. We note that the signal from the balanced photodetector is proportional to a Faraday rotation angle times laser power $P$. So, in what follows we show the Faraday rotation signal obtained by normalizing the measured signal to $P$. 

The rf magnetic field is applied along the sample normal using a small coil ($1$-mm-inner and $1.5$-mm-outer diameter) near the sample surface. Current through the coil is driven by a function generator, which creates a sinusoidal voltage with a frequency $f_\text{rf}$ up to 150~MHz and an amplitude $U_\text{rf}$ up to 10~V. The amplitude of the rf field $b$ is proportional to $U_\text{rf}$: $b = k U_\text{rf}/f_\text{rf}$. The coefficient $k$ at high $f_\text{rf}$ can be estimated as $k = 1/(2\pi^2 N r^2) \approx 10$~mT MHz$/$V, where $r\approx 0.7$~mm is the coil radius, $N = 10$ is the number of windings \cite{Belykh2019RPOP}. The generator output is modulated at a frequency $f_\text{m}$ for synchronous detection with a lock-in amplifier. In order to perform the spin inertia method, the modulation frequency $f_\text{m}$ was varied. The measured signal is the difference between the Faraday rotation values for the low and high levels of the rf field, which is in turn proportional to the corresponding difference $\Delta S$ in the spin polarizations~\cite{Belykh2020,BelykhPRL2021}. 

\section{ODMR spectra and $g$-factor anisotropy}
\begin{figure*}
\includegraphics[width=2\columnwidth]{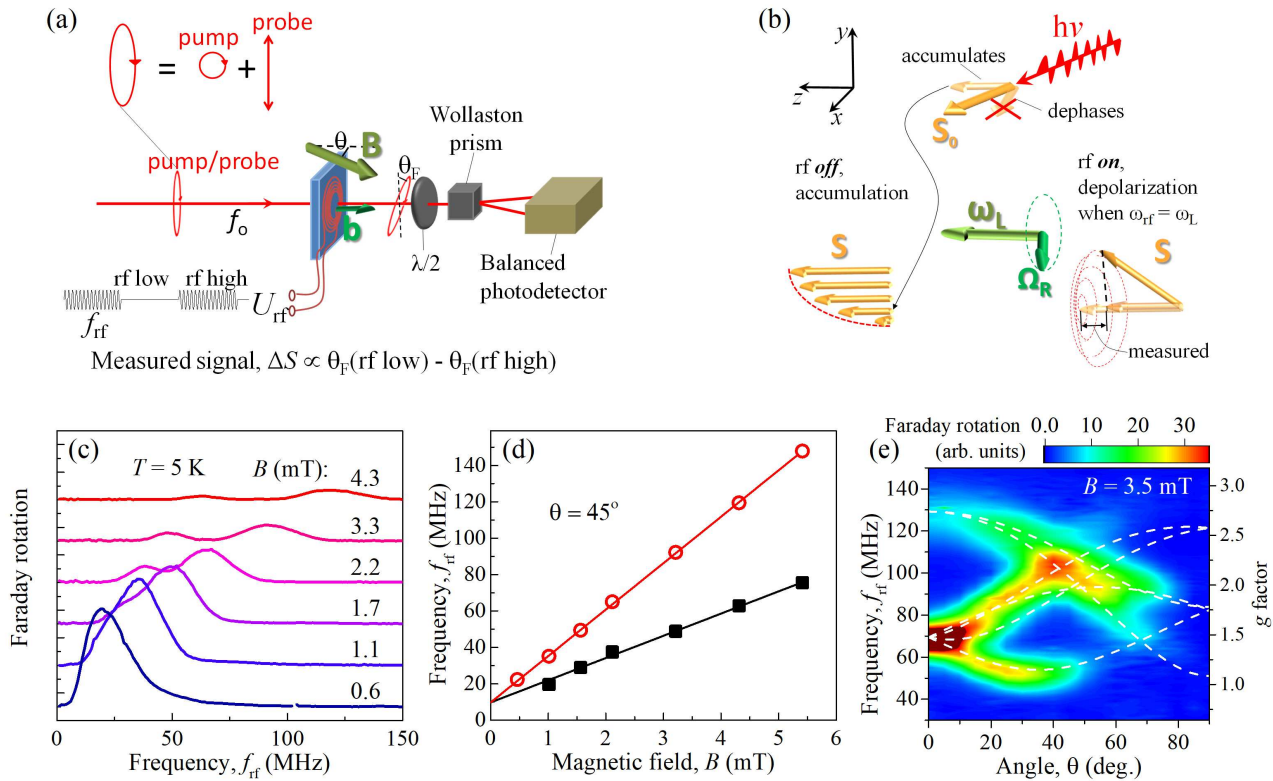}
\caption{(a) Scheme of the experiment. (b) Qualitative interpretation of the effect: optical pumping leads to the accumulation of spin polarization along the Larmor frequency $\boldsymbol{\omega}_\text{L}$, while the rf field resonant with the Larmor frequency unwinds spin polarization from the direction of $\boldsymbol{\omega}_\text{L}$ and suppresses spin accumulation. (c) Faraday rotation signal as a function of the rf field frequency (ODMR spectra) for different values of the magnetic field applied at the angle $\theta = 45^\circ$ with respect to the sample normal. (d) Magnetic field dependences of the frequencies corresponding to the two main peaks in the ODMR spectra. Their slopes give $g$ factor values of 0.9 and 1.9. (e) Faraday rotation signal (shown by color) as a function of the angle of the magnetic field with respect to the sample normal and the rf field frequency. Left axis shows the $g$ factor values. Dashed lines show calculation results as described in Appendix A. $B = 3.5$~mT. (c)-(e) $T = 5$~K.}
\label{fig:BDep}
\end{figure*}

The Ce$^{3+}$ ion has one unpaired electron in the 4f level, which can be excited optically to the 5d level. The 4f and 5d levels contain 7 and 5 Kramers doublets (twofold spin degenerate states), respectively, which are split by the spin-orbit coupling and crystal field. In our experiment we excite electron from the lowest in energy doublet in 4f level to the lowest doublet in 5d level in a phonon-assisted absorption process, which has a relatively broad band \cite{Robbins1979} fitting the used wavelength of 444~nm. Circularly polarized light excites electrons with a certain spin (spin-down in the case of $\sigma^+$ polarization), which is flipped in the course of excitation. Meanwhile, upon their relaxation back to the ground 4f level, electrons may end up in spin-down or spin-up state with about the same probabilities \cite{Kolesov2013,Siyushev2014}. In this way, the electrons occupying the ground 4f level in the ensemble of Ce$^{3+}$ ions become preferentially spin-up polarized under $\sigma^+$ excitation. The characteristic time of electron relaxation from 5d to 4f level is about 70~ns \cite{Zych2000}. It is much shorter than the microsecond-long spin dynamics considered here. Thus, the latter is not contributed by the excited 5d states. The energy level structure of the Ce$^{3+}$ ion and the scheme of its optical orientation is described in detail in Refs.~\cite{Kolesov2013,Siyushev2014,Azamat2017,Liang2017}. 

As a result of optical orientation, the electron spin ensemble acquires spin polarization $\mathbf{S}$ along the laser beam. The transverse component of $\mathbf{S}$, perpendicular to the Larmor frequency $\boldsymbol{\omega}_\text{L} = (\mu_\text{B}/\hbar)\hat{g} \mathbf{B}$ (note that in general $\boldsymbol{\omega}_\text{L} \nparallel \mathbf{B}$ for anisotropic $g$ tensor $\hat{g}$), precesses around $\boldsymbol{\omega}_\text{L}$ and becomes dephased rapidly on a timescale of the inhomogeneous dephasing time $T_2^* \sim 10$~ns \cite{Azamat2017}, unless the laser pulse repetition rate is coherent with the Larmor frequency. The coherent case was considered in Ref.~\onlinecite{BelykhPRL2021} and will not be discussed here. On the other hand, the longitudinal (along $\boldsymbol{\omega}_\text{L}$) component of $\mathbf{S}$ lives much longer, it decays with the time $T_1$. As a result, spin polarization accumulates along $\boldsymbol{\omega}_\text{L}$ [Fig.~\ref{fig:BDep}(b)] and saturates at a level proportional to the pump power $P$ and $T_1$. When an rf field is applied, it reduces the efficiency of the spin accumulation by unwinding spin from the $\boldsymbol{\omega}_\text{L}$ direction, which, as is shown below, can be described as additional relaxation. The effect of the rf field is maximal when its frequency $f_\text{rf}$ is resonant with the Larmor frequency $f_\text{L} = \omega_\text{L}/2\pi$. This is illustrated in Fig.~\ref{fig:BDep}(c), which shows the rf field frequency dependence of the Faraday rotation signal, proportional to the difference of spin polarizations at the low and high rf field levels. The spectra show pronounced peaks when $f_\text{rf}$ coincides with $f_\text{L}$; this behavior resembles optically detected magnetic resonance (ODMR). The peak positions depend on the permanent magnetic field $B$, which is summarized in Fig.~\ref{fig:BDep}(d). The dependence is linear in accordance with the equation $f_\text{L} = |g|\mu_\text{B}B/2\pi\hbar$. The small offset of 10~MHz at $B=0$ is presumably related to internal nuclear magnetic fields present in the system. The two peaks correspond to $|g| = 0.9$ and $|g| = 1.9$. Different $g$ factors in the Ce$^{3+}$:YAG system arise from the six different orientations of the highly anisotropic $g$ tensor of Ce$^{3+}$ ions embedded in the six different $c$ sites in the YAG lattice. The $g$ tensor is characterized by the main values of 2.738, 1.872 and 0.91 along 3 perpendicular axes \cite{Lewis1966}. Indeed, the positions of the peaks in the ODMR spectra show strong dependence on the angle $\theta$ of the magnetic field with respect to the sample normal [Fig.~\ref{fig:BDep}(e)]. The description of the angular dependence of the observed resonances is given in Appendix A. Here, it is important that we can observe and resonantly address specific Larmor frequencies. Note that the signal generally decreases with angle $\theta$ when going from the Faraday to Voigt geometry. Indeed, the spin polarization is created along the sample normal and its component along $\boldsymbol{\omega}_\text{L}$, which is accumulated, decreases with $\theta$. The detected signal is in turn proportional to the component of the accumulated spin along the sample normal, which also decreases with $\theta$.

\section{Longitudinal spin relaxation}
\begin{figure*}
\includegraphics[width=1.5\columnwidth]{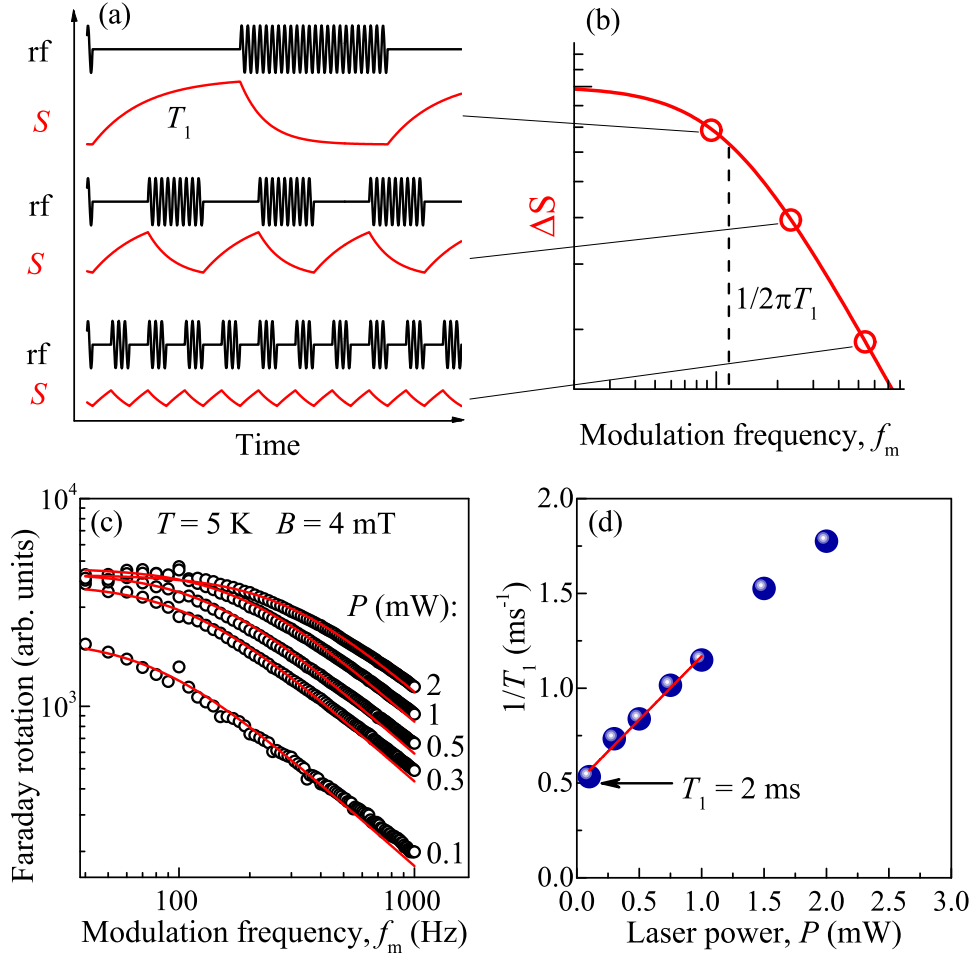}
\caption{(a) rf field protocols with different modulation frequencies and corresponding calculated spin polarization dynamics. When the rf modulation period becomes smaller than $T_1$, the modulation depth $\Delta S$ of the spin polarization is reduced. (b) Calculated spin polarization amplitude $\Delta S$ as a function of rf field modulation frequency (spin inertia curve). (c) Faraday rotation signal as a function of the rf field modulation frequency (spin inertia curves) for different laser powers. Red lines show the fits to the experimental data with Eq.~\eqref{eq:RSmallW}. (d) Longitudinal spin relaxation rate $1/T_1$ extracted from the spin inertia curves as a function of the laser power. Red line shows the linear fit which is used to extrapolate the dependence to $P = 0$. (c)-(d) $B = 4$~mT, $\theta = 12^\circ$, $f_\text{rf} = 80$~MHz, and $T = 5$~K.}
\label{fig:PDep}
\end{figure*}

In the absence of the rf field, spin polarization saturates at the level proportional to the spin relaxation time $T_1$ and the pumping rate $P$. When the rf field is turned \textit{on}, spin polarization begins to decrease and saturates at some lower level. The measured signal $\Delta S$ is given by the difference between these saturation levels. This remains true while the rf modulation period is much longer than the time $T_1$, so that spin polarization has time to accumulate when the rf field is {\it off} and have time to decay when the rf field is {\it on}. When the frequency $f_\text{m}$ of the rf modulation is increased, so that the modulation period $T_\text{m} = 1/f_\text{m}$ becomes shorter than $T_1$, spin polarization accumulated in the first half of the modulation period becomes smaller, and $\Delta S$ decreases. This fact is illustrated in Fig.~\ref{fig:PDep}(a) and can be used to determine $T_1$. Indeed, if one measures the dependence of $\Delta S$ on $f_\text{m}$, the signal remains constant at low $f_\text{m}$, while for $f_\text{m} \gg 1/T_1$ the signal is proportional to $1/f_\text{m}$ [Fig.~\ref{fig:PDep}(b)]. The crossover frequency in this dependence gives $1/T_1$. This comprises the spin inertia method. The spin-inertia principle was introduced all-optically with separate spin pumping and probing and varying the modulation frequency of the pump beam \cite{Heisterkamp2015}. In our case it is the spin relaxation rate rather than optical pumping that is modulated, which allows us to tune the rf field frequency and, therefore, to measure $T_1$ for a particular resonance. For the small amplitude of the rf field used in most of our experiments, the dependence of  $\Delta S$ on $f_\text{m}$ can be described by the spin inertia equation~\eqref{eq:RSmallW}, as shown below.

The experimental dependences of the Faraday rotation signal on the rf field modulation frequency are shown in Fig.~\ref{fig:PDep}(c) for different pump powers, magnetic field of $4$~mT tilted by $\theta = 12^\circ$ with respect to the sample normal, and $f_\text{rf} = 80$~MHz, corresponding to $|g| = 1.4$. The spike at 100~Hz corresponds to the equipment-based resonance. The values of $T_1$ can be determined by fitting these dependences with Eq.~\eqref{eq:RSmallW}, while an even more rigorous way to determine $T_1$ involving signal retardation is described in the theory section. The time $T_1$ turns out to be few ms for this system in agreement with Refs~\cite{Siyushev2014,Azamat2017}. The measured spin relaxation rate $1/T_1$ expectedly increases with the laser power [Fig.~\ref{fig:PDep}(d)], because, apart from creating spin polarization, the laser beam also disturbs the already created polarization. A linear extrapolation of $1/T_1$ to the limit of $P = 0$ allows one to determine the value $T_1 \approx 2$~ms for the unpumped system. This value is close to $T_1$ measured at $B = 49$~mT and $T = 3.5$~K for a single Ce$^{3+}$ ion in YAG lattice in Ref.~\onlinecite{Siyushev2014}.

\section{Temperature dependence of $T_1$}
\begin{figure}
\includegraphics[width=0.8\columnwidth]{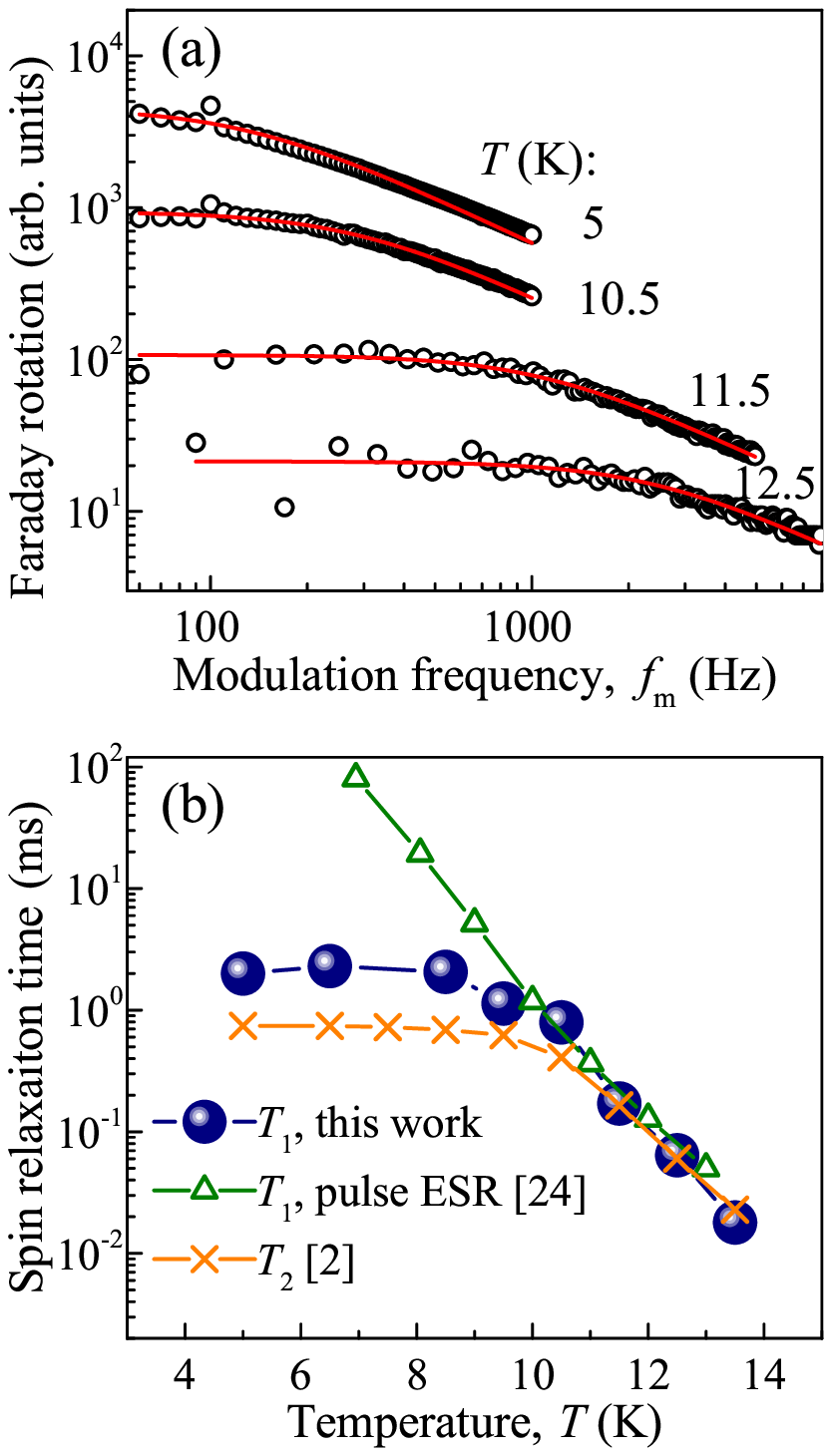}
\caption{(a) Spin inertia curves for different temperatures. Red lines show the fit to the experimental data with Eq.~\eqref{eq:RSmallW}. $B = 4$~mT, $\theta = 12^\circ$, $f_\text{rf} = 80$~MHz. (b) Summary of the temperature dependences of $T_1$ from this work at $B = 4$~mT (balls) in the limit of $P = 0$, from Ref.~\onlinecite{Azamat2017} at $B = 474.4$~mT (triangles), and $T_2$ from Ref.~\onlinecite{BelykhPRL2021} at $B = 3.3$~mT (crosses) for Ce$^{3+}$:YAG.}
\label{fig:TDep}
\end{figure}

An increase in temperature from 5 to 13~K expectedly results in the shortening of $T_1$. This manifests itself as a broadening of the frequency range where the signal is constant in the spin inertia curves [Fig.~\ref{fig:TDep}(a)]. The temperature dependence of $T_1$ is compared to that measured by the pulse-ESR technique in Ref.~\onlinecite{Azamat2017} and to the temperature dependence of $T_2$ measured by stimulated resonant spin amplification in Ref.~\onlinecite{BelykhPRL2021} [Fig.~\ref{fig:TDep}(b)]. Note that pulse-ESR reveals two components of the longitudinal spin relaxation, while here we show only the slow component [triangles in Fig.~\ref{fig:TDep}(b)]. At $T > 10$~K all the dependences show the same behavior, which can be described by the combination of the two-phonon Raman process and the activation with a longitudinal optical (LO) phonon \cite{Azamat2017,BelykhPRL2021}. When temperature is decreased, $T_1$ measured here and $T_2$ from Ref.~\onlinecite{BelykhPRL2021} saturate, while time $T_1$ from Ref.~\onlinecite{Azamat2017} continue to increase steadily. This difference might be related to the pronouncedly different magnetic fields at which measurements were performed: $B = 474.4$~mT for the pulse-ESR and few mT for this work and for $T_2$ measurements in Ref.~\onlinecite{BelykhPRL2021}. Indeed, an external magnetic field suppresses the effect of fluctuating nuclear fields on the electron spin relaxation \cite{Merkulov2002}.

\section{Magnetic field dependence and anisotropy of $T_1$}
\begin{figure}
\includegraphics[width=0.8\columnwidth]{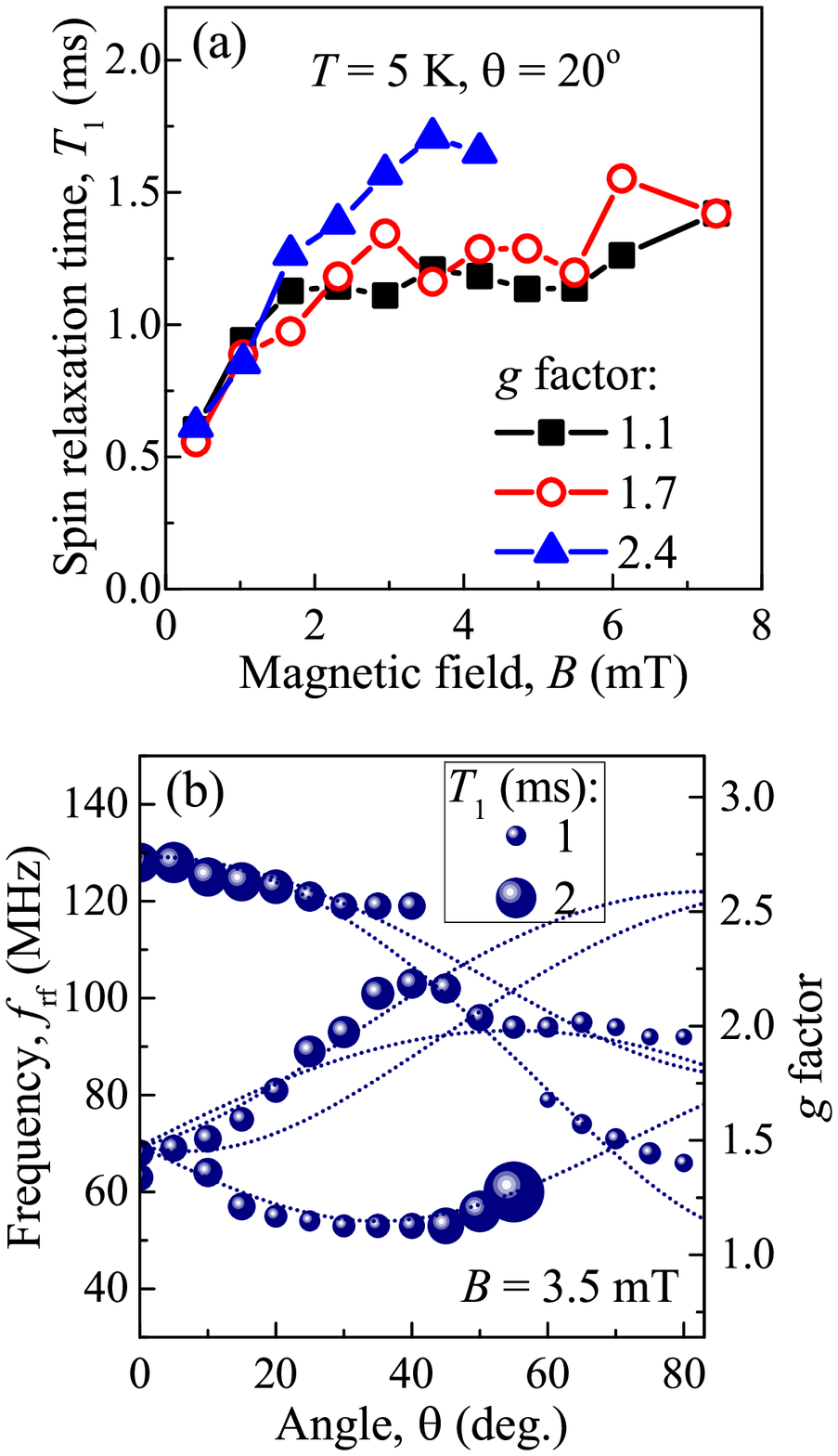}
\caption{(a) Magnetic field dependences of $T_1$ for three different $g$ factors and $\theta = 20^\circ$. (b) Values of $T_1$, reflected by the sizes of the bubbles, as a function of the magnetic field angle $\theta$ and the rf field frequency $f_\text{rf}$. Left axis shows the $g$ factor values. Dotted lines show calculations as described in Appendix A. $B = 3.5$~mT. (a),(b) $P = 0.5$~mW, $T = 5$~K.}
\label{fig:T1Anisotr}
\end{figure}
The magnetic field dependences of $T_1$ for a fixed angle $\theta = 20^\circ$ and three different $g$ factors are shown in Fig.~\ref{fig:T1Anisotr}(a). The time $T_1$ increases when $B$ is increased up to 3~mT and then saturates. The values of $T_1$ for $B \gtrsim 3$~mT are different for different $g$ factors. One may guess that $T_1$ increases with $g$, but the situation is actually much more complicated. We measure the anisotropy of $T_1$ by varying the magnetic field angle $\theta$ and the frequency $f_\text{rf}$ and following the spin resonance angular dependences shown in Fig.~\ref{fig:BDep}(e). The values of $T_1$ along these dependences are reflected by the sizes of the bubbles in Fig.~\ref{fig:T1Anisotr}(b). It follows from this figure that there is no unique correspondence between $T_1$ and $g$ factor, while $T_1$ strongly depends on the orientation of the magnetic field and can change more than a factor of 2 with $\theta$. This strong anisotropy of $T_1$ is presumably related to the fact that under the conditions of the low magnetic fields and low temperatures $T_1$ is temperature-independent and is determined by the electron-nuclear spin interaction having anisotropic nature. 

\section{Theory}
In this section, we consider the effect of the rf field on an inhomogeneous spin system and give a detailed description of the resonant spin inertia effect.
\subsection{General description}
The evolution of the electron spin in an external magnetic field under optical pumping can be described by the Bloch equation \cite{Bloch1946}:
\begin{equation}
\frac{d\mathbf{S}}{dt} = (\boldsymbol{\omega}_\text{L} + \boldsymbol{\Omega}_\text{R}(t)) \times \mathbf{S} - \hat{\gamma}\mathbf{S} + \mathbf{P},
\label{eq:RPOP:BlochW}
\end{equation}
where $\boldsymbol{\Omega}_\text{R}(t) = \Omega_\text{R} \begin{pmatrix}
\cos(\omega_\text{rf}t)\\
\sin(\omega_\text{rf}t)\\
0
\end{pmatrix}$, $\Omega_\text{R} = (\mu_\text{B}/2\hbar) |\hat{g}\mathbf{b}|\sin(\theta')$ is the Rabi frequency, $\theta'$ is the angle between $\boldsymbol{\omega}_\text{L}$ and $\hat{g}\mathbf{b}$, $\mathbf{b}$ is the amplitude of the rf field, $\mathbf{P}$ is the pumping rate, and $\hat{\gamma} = \begin{pmatrix} 1/T_2 & 0 & 0\\ 0 & 1/T_2 & 0\\ 0 & 0 & 1/T_1 \end{pmatrix}$ is the relaxation matrix. Here we choose the coordinate system with $z$-axis along $\boldsymbol{\omega}_\text{L}$, so that $\mathbf{P}$ (laser beam) and $\mathbf{b}$ lay in the $xz$ plane.
We take into account only the transverse component of the rf field [more precisely of $\hat{g}\mathbf{b}(t)$] with respect to $\boldsymbol{\omega}_\text{L}$ which is then decomposed into two circular components rotating in the opposite directions with the frequencies $\boldsymbol{\omega}_\text{rf}$ and $-\boldsymbol{\omega}_\text{rf}$. Only the component with the frequency $\boldsymbol{\omega}_\text{rf}$ directed along $\boldsymbol{\omega}_\text{L}$ is finally left in the rotating wave approximation as the other one is strongly out of resonance \cite{Abragam1961}. One can show that the accumulated spin polarization is mainly determined by the longitudinal (with respect to $\boldsymbol{\omega}_\text{L}$) component of $\mathbf{P}$, while the effect of the transverse component is of the order of $1/\omega_\text{L}T_2 \ll 1$. So to simplify the equations we assume that $\mathbf{P} \parallel \boldsymbol{\omega}_\text{L}$ implying that $\mathbf{P}$ is the longitudinal component of the pumping rate.  

Without the rf field, $\boldsymbol{\Omega}_\text{R}(t) = 0$, the solution of Eq.~\eqref{eq:RPOP:BlochW} is 
\begin{multline}
\mathbf{S}(t) = \begin{pmatrix}
\left[S_x(0) \cos(\omega_\text{L}t)-S_y(0) \sin(\omega_\text{L}t)\right] \exp(-t/T_2)\\
\left[S_y(0) \cos(\omega_\text{L}t)+S_x(0) \sin(\omega_\text{L}t)\right] \exp(-t/T_2)\\
(S_z(0) - P T_1) \exp(-t/T_1)
\end{pmatrix} \\ +
\begin{pmatrix}
0\\
0\\
P T_1
\end{pmatrix}.
\end{multline}
Note that $\mathbf{S}(t)$ tends to its steady-state value 
\begin{equation}
\mathbf{S}_\text{st} = P T_1 \begin{pmatrix}
0\\0\\1
\end{pmatrix}.
\end{equation}
In our experiments, oscillations on the timescale of $1/\omega_\text{L}$ are averaged out and the relevant spin dynamics is left only in the $z$ component of $\mathbf{S}$, along $\boldsymbol{\omega}_\text{L}$:
\begin{equation}
S_z(t) = S_z(0)\exp(-t/T_1) +P T_1 [1-\exp(-t/T_1)]. 
\label{eq:SnoRF}
\end{equation}

When the rf field is \textit{on}, we can change to the reference frame rotating with the frequency $\boldsymbol{\omega}_\text{rf}$ directed along $\boldsymbol{\omega}_\text{L}$ by changing the variables:
\begin{eqnarray}
\mathbf{S} = \hat{R}\mathbf{S}',\\
\hat{R} = 
\begin{pmatrix}
\cos(\omega_\text{rf}t) & -\sin(\omega_\text{rf}t) & 0 \\
\sin(\omega_\text{rf}t) & \cos(\omega_\text{rf}t) & 0 \\
0 & 0 & 1
\end{pmatrix}.
\label{eq:RS}
\end{eqnarray}
Then the Bloch equation reads as
\begin{equation}
\frac{d\mathbf{S}'}{dt} = \boldsymbol{\Omega} \times \mathbf{S}' - \hat{\gamma}\mathbf{S}' + \mathbf{P},
\label{eq:RPOP:BlochRF}
\end{equation}
where $\boldsymbol{\Omega} = \begin{pmatrix} \Omega_\text{R}\\ 0\\ \Delta\omega_\text{L}\end{pmatrix}$, $\Delta\omega_\text{L} = \omega_\text{L} - \omega_\text{rf}$. The stationary value of the spin polarization in this case is 
\begin{equation}
\mathbf{S}_\text{st}' = \frac{P T_1}{1+\Delta\omega_\text{L}^2 T_2^2 + \Omega_\text{R}^2 T_2 T_1} 
\begin{pmatrix}
\Omega_\text{R} \Delta\omega_\text{L} T_2^2\\ 
-\Omega_\text{R} T_2 \\ 
1+\Delta\omega_\text{L}^2 T_2^2 \end{pmatrix}.
\end{equation}
Equation~\eqref{eq:RPOP:BlochRF} can be solved analytically, however its solution is too cumbersome. To simplify it we assume that $|\Delta\omega_\text{L}|, \Omega_\text{R} \gg 1/T_1, 1/T_2$. With this approximation $\mathbf{S}_\text{st}' \approx \mathbf{\Omega} P T_1 \Delta\omega_\text{L}/(\Delta\omega_\text{L}^2 + \Omega_\text{R}^2 T_1/T_2)$ and the solution of Eq.~\eqref{eq:RPOP:BlochRF} reads as:
\begin{multline}
\mathbf{S}' = \mathbf{S}_\text{st}'[1-\exp(-t/\tau_1)] + (\mathbf{S}'(0) \mathbf{n})\mathbf{n}\exp(-t/\tau_1)\\
+ [\mathbf{S}'(0) -(\mathbf{S}'(0) \mathbf{n})\mathbf{n}]\cos(\Omega t)\exp(-t/\tau_2)\\
+\mathbf{n}\times\mathbf{S}'(0)\sin(\Omega t)\exp(-t/\tau_2),
\end{multline}
where $\mathbf{n} = \mathbf{\Omega}/\Omega$, 
\begin{eqnarray}
\tau_1^{-1} = \frac{\Delta\omega_\text{L}^2}{\Omega^2}T_1^{-1}+\frac{\Omega_\text{R}^2}{\Omega^2}T_2^{-1},\\
\tau_2^{-1} = \frac{\Omega_\text{R}^2}{2\Omega^2}T_1^{-1}+(1-\frac{\Omega_\text{R}^2}{2\Omega^2})T_2^{-1}.
\end{eqnarray}
Thus, in the rotating reference frame the longitudinal component of the spin with respect to $\mathbf{\Omega}$ relaxes to $\mathbf{S}_\text{st}'$ with the time $\tau_1$, while the transverse component precesses and decays with the time $\tau_2$. When changing back to the laboratory reference frame, the $x$ and $y$ spin components acquire oscillating factors $\cos(\omega_\text{rf}t)$ and $\sin(\omega_\text{rf}t)$ which are averaged out to zero in our experiment. So we again are interested only in $S_z(t)=S_z'(t)$:
\begin{multline}
S_z(t) = \frac{P T_1}{1 + \Omega_\text{R}^2 T_1/\Delta\omega_\text{L}^2 T_2}[1-\exp(-t/\tau_1)] \\
+ S_z(0)\frac{\Delta\omega_\text{L}^2}{\Omega^2}\exp(-t/\tau_1) \\
+ S_z(0)\frac{\Omega_\text{R}^2}{\Omega^2}\cos(\Omega t)\exp(-t/\tau_2),
\label{eq:SxRF}
\end{multline}
where we assume that $S_x(0) = S_y(0) = 0$ due to averaging. The third term of the right-hand side of the Eq.~\eqref{eq:SxRF} describes the rapid (on the timescale of $1/\Omega$) decrease of the spin polarization right after switching the rf field {\it on}. The amount of the residual long-living spin polarization is described by the factor $\Delta\omega_\text{L}^2/\Omega^2=1/(1+\Omega_\text{R}^2/\Delta\omega_\text{L}^2)$ in the second term. Thus, the effect of the rf field depends on how far is the Larmor frequency from the rf field frequency. The detuning $\Delta\omega_\text{L}$ may have two origins. (i) Spread of the Larmor frequencies in the electron spin ensemble due to inhomogeneous frozen internal magnetic fields or spread of $g$ factors. (ii) Temporal fluctuations of $\omega_\text{L}$ for each electron in a time-varying environment. In both cases the maximal variation of $\Delta\omega_\text{L}$ can be estimated from the half-width at half-maximum (HWHM) of the peaks in ODMR spectra [Fig.~\ref{fig:BDep}(c)], $\Delta\omega_\text{max} \sim 2\pi\times$ (10~MHz).

In further calculations we take into account the following. In the first half of the rf field modulation period, when the rf field is {\it off}, spin dynamics is described by Eq.~\eqref{eq:SnoRF}, while in the second half-period, when the rf field is \textit{on}, it is described by Eq.~\eqref{eq:SxRF}. The initial conditions for both equations can be found from the continuity requirement at times 0, $T_\text{m}/2$, and $T_\text{m}$.  

\begin{figure*}
\includegraphics[width=1.5\columnwidth]{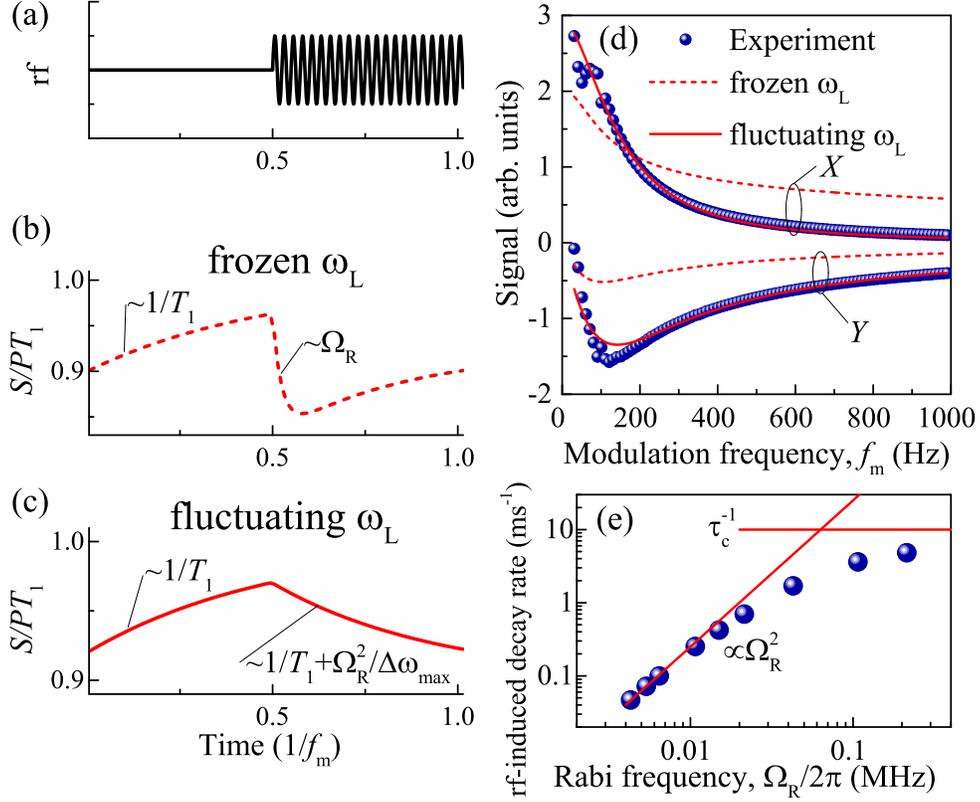}
\caption{(a) rf field profile. (b) Schematic representation of the ensemble spin dynamics for frozen Larmor frequencies. (c) Schematic representation of the ensemble spin dynamics for time-fluctuating Larmor frequencies. (d) Frequency dependence of the spin dynamics convolved with sine [$X$, $\sin(2\pi f_\text{m}t)$] and cosine [$Y$, $\cos(2\pi f_\text{m}t)$] functions. Dashed and solid lines show calculations for the assumptions of frozen and time-fluctuating $\omega_\text{L}$. Time-fluctuating $\omega_\text{L}$ results in additional decay of the spin polarization when the rf field is \textit{on}. The rate of this decay as a function of the Rabi frequency (rf field amplitude) is shown in panel (e).(d)-(e) $B = 4$~mT, $\theta = 12^\circ$, $f_\text{rf} = 80$~MHz, $P = 1$~mW, $T = 5$~K.}
\label{fig:EnsTime}
\end{figure*}
\subsection{Frozen magnetic field fluctuations}
For the case of frozen magnetic field fluctuations one has to average Eqs.~\eqref{eq:SnoRF},\eqref{eq:SxRF} over $\Delta\omega_\text{L}$ distribution centered at zero and having HWHM of $\Delta\omega_\text{max}$. We demonstrate the result of this averaging for $T_1=T_2$ and assume a Lorentzian distribution of the Larmor frequencies: $\Delta\omega_\text{max}/\pi(\omega_\text{max}^2+\Delta\omega_\text{L}^2)$. The averaged spin dynamics is given by Eq.~\eqref{eq:SxRFEns} in Appendix B and schematically illustrated in Fig.~\ref{fig:EnsTime}(b). In the first half-period of the rf modulation, the spin polarization exponentially tends to the stationary value $P T_1$. Next, switching the rf field {\it on} results in a rapid decrease of the spin polarization, described by the factor $\Delta\omega_\text{max}/(\Delta\omega_\text{max}+\Omega_\text{R})$. During the second half-period the reduced spin polarization tends to the stationary value reduced by the same factor [Fig.~\ref{fig:EnsTime}(b)]. In the experiment, the synchronous detection scheme is performed and the lock-in amplifier measures the convolution of the spin signal with $\sin(2\pi f_\text{m} t)$ in the $X$ channel and $\cos(2\pi f_\text{m} t)$ in the $Y$ channel. The results of such convolutions for Eq.~\eqref{eq:SxRFEns} are given by Eqs.~\eqref{eq:Xfroz},\eqref{eq:Yfroz}. For $\Omega_\text{R} \ll \Delta\omega_\text{max}$, which is the case for most of our experiments, Eqs.~\eqref{eq:Xfroz}\eqref{eq:Yfroz} reduce to 
\begin{multline}
X = \frac{P T_1}{\pi\Delta\omega_\text{max}}\Omega_\text{R}  \Big[1 - \frac{2\pi^2 f_\text{m}^2 T_1^2}{1 + 4 \pi^2 T_1^2 f_\text{m}^2}\\
\times\left\{1 + \exp\left(\frac{1}{2T_1 f_\text{m}}\right) - \sqrt{\exp\left(\frac{1}{T_1 f_\text{m}}\right)-1}\right\} \Big]
\label{eq:XfrozSmall}
\end{multline}
\begin{multline}
Y = - \frac{P T_1}{\Delta\omega_\text{max}}\Omega_\text{R} \frac{f_\text{m} T_1}{1 + 4 \pi^2 T_1^2 f_\text{m}^2}\\
\times\left\{1 + \exp\left(\frac{1}{2T_1 f_\text{m}}\right) - \sqrt{\exp\left(\frac{1}{T_1 f_\text{m}}\right)-1}\right\}
\label{eq:YfrozSmall}
\end{multline}

The experimental dependence of the signals in the $X$ and $Y$ channels of the lock-in amplifier are shown in Fig.~\ref{fig:EnsTime}(d). The best fit based on the model of the frozen $\omega_\text{L}$, with Eqs.~\eqref{eq:XfrozSmall},\eqref{eq:YfrozSmall}, shown by the dashed lines in Fig.~\ref{fig:EnsTime}(d), is far from the experimental dependences. We have also checked by numerical calculations that taking into account that $T_1 \neq T_2$ [Fig.~\ref{fig:TDep}(b)] does not improve the fit much. 

\subsection{Time-varying magnetic field fluctuations}
Next, we take into account temporal fluctuations of $\omega_\text{L}$, which apparently originate from spin fluctuations of the surrounding nuclei. Here, apart from the amplitude of the Larmor frequency fluctuations $\Delta\omega_\text{L}$, we have to introduce the correlation time $\tau_\text{c}$ of these fluctuations, so that the Larmor frequency changes by $\sim \Delta\omega_\text{max}$ over a time period $t \gtrsim \tau_\text{c}$. Let us estimate the change of $S_z$ during time $\tau_\text{c}$. We assume $\Omega_\text{R} \ll \Delta\omega_\text{max}$. One can see from Eq.~\eqref{eq:SxRF} that the rf field barely affects the spin polarization when $|\Delta\omega_\text{L}| \sim \Delta\omega_\text{max}$. However, when $\Delta\omega_\text{L}$ becomes close to zero, and comparable to $\Omega_\text{R}$, $S_z$ is reduced with the rate determined by the last term in Eq.~\eqref{eq:SxRF}. It shows that the lower is $\Delta\omega_\text{L}$, the larger is the change in $S_z$ ($\Omega_\text{R}^2/\Omega^2$), but the longer is the time required for this change to occur ($1/\Omega$). The time interval where $|\Delta\omega_\text{L}| \lesssim \Omega$ can be estimated as $\Delta t \sim \tau_\text{c}\Omega/\Delta\omega_\text{max}$. Let us find $\Omega$ for which this time is long enough, $\Delta t \sim 1/\Omega$, so that the last term in Eq.~\eqref{eq:SxRF} changes from the maximum to zero: $\Omega^2 \sim \Delta\omega_\text{max} / \tau_\text{c}$. The change of the spin polarization in time $\tau_\text{c}$ can be estimated as $\Delta S_z/S_z \sim -\Omega_\text{R}^2/\Omega^2 \sim -\tau_\text{c}\Omega_\text{R}^2/\Delta\omega_\text{max}$. Thus, the rf-induced decrease of the spin polarization $dS_z/dt \approx \Delta S_z/\tau_\text{c} \sim - (\Omega_\text{R}^2/\Delta\omega_\text{max})S_z$ is exponential and can be described by the time $\tau_\text{rf} = \Delta\omega_\text{max} / \Omega_\text{R}^2$. One can note that if  $\Omega_\text{R}^2 \gtrsim \Delta\omega_\text{max} / \tau_\text{c}$, the relative change of the spin polarization is about 1 and $\tau_\text{rf} \sim \tau_\text{c}$. In this way, the dynamics of the spin polarization averaged over many periods of the rf modulation for the time fluctuating Larmor frequency can be described by the equation
\begin{multline}
<S_z>_\text{time}(t) \approx S_z (0) \exp(-t/T_1-t/\tau_\text{rf}) \\
+ \frac{P}{1/T_1+1/\tau_\text{rf}}[1-\exp(-t/T_1-t/\tau_\text{rf})],
\end{multline}
where 
\begin{equation}
\tau_\text{rf}^{-1} \sim \begin{cases}
   	\Omega_\text{R}^2/\Delta\omega_\text{max}, \Omega_\text{R}^2 \ll \Delta\omega_\text{max} / \tau_\text{c};\\
	\tau_\text{c}^{-1},  \Omega_\text{R}^2 \gtrsim \Delta\omega_\text{max} / \tau_\text{c}.
	\end{cases}
	\label{eq:taurf}
\end{equation}
In this way, the application of the rf field modulates the relaxation rate. The temporal profile of the averaged spin polarization is schematically illustrated in Fig.~\ref{fig:EnsTime}(c) and is given by Eqs.~\eqref{eq:SxRFTime} in Appendix C. In the first half-period the spin polarization increases towards the stationary value of $PT_1$ with the rate of $1/T_1$, while in the second half-period the spin polarization decreases towards $P/(T_1^{-1}+\tau_\text{rf}^{-1})$ with the rate of $1/T_1+1/\tau_\text{rf}$. The convolution of this spin temporal profile with $\sin(2\pi f_\text{m} t)$ and $\cos(2\pi f_\text{m} t)$ provides the signals measured in the $X$ and $Y$ channels, respectively, of the lock-in amplifier, which are given by Eqs.~\eqref{eq:X},\eqref{eq:Y} in Appendix C. These equations provide a good fit to the measured signal, proportional to the Faraday rotation, in the $X$ and $Y$ channels of the lock-in amplifier which is shown in Fig.~\ref{fig:EnsTime}(d) by the red solid lines. From this fit, one can determine $T_1$ and $\tau_\text{rf}$. By fitting the data corresponding to different voltages applied to the rf coil, i.e., to different $\Omega_\text{R}$ (obtained from the values of $b$, see the experimental details section), we plot the dependence of $\tau_\text{rf}^{-1}$ on $\Omega_\text{R}$ [Fig.~\ref{fig:EnsTime}(e)]. The result is in good agreement with the estimate of Eq.~\eqref{eq:taurf}: at small rf fields, $\tau_\text{rf}^{-1}$ is quadratic in $\Omega_\text{R}$, and then levels off at $\tau_\text{c}^{-1}$. From this dependence, we can estimate $\tau_\text{c} \sim 0.1$~ms. 

Interestingly, for the small rf-induced relaxation rate, $\tau_\text{rf}^{-1} \ll T_1^{-1}$ ($\Omega_\text{R}^2 \ll \Delta\omega_\text{max}/T_1$), which is the case in most of our experiments, Eqs.~\eqref{eq:X},\eqref{eq:Y} reduce to 
\begin{eqnarray}
X \approx \frac{P T_1^2}{\pi\Delta\omega_\text{max}}\Omega_\text{R}^2\frac{1}{1+4\pi^2 T_1^2 f_\text{m}^2}
\label{eq:XSmallW}\\
Y \approx -\frac{P T_1^2}{\pi\Delta\omega_\text{max}}\Omega_\text{R}^2\frac{2\pi T_1 f_\text{m}}{1+4\pi^2 T_1^2 f_\text{m}^2}
\label{eq:YSmallW}
\end{eqnarray}
Note, these equations also hold for the sinusoidal modulation of the rf field.
The magnitude of the Faraday rotation signal (spin polarization), which is plotted in Figs.~\ref{fig:PDep}(c) and \ref{fig:TDep}(a), is proportional to $\sqrt{X^2+Y^2}$:
\begin{equation}
\Delta S \propto \frac{P T_1^2}{\pi\Delta\omega_\text{max}}\Omega_\text{R}^2\frac{1}{\sqrt{1+4\pi^2 T_1^2 f_\text{m}^2}}
\label{eq:RSmallW}
\end{equation}
which resembles the classical spin inertia equation \cite{Heisterkamp2015,Zhukov2018,Smirnov2018} apart from the prefactor. 
Nevertheless, since in the experiments it is possible to measure signals in $X$ and $Y$ channel separately, we use Eqs.~\eqref{eq:XSmallW},\eqref{eq:YSmallW} to determine $T_1$ by fitting two sets of data. This expectedly provides a better accuracy than Eq.~\eqref{eq:RSmallW} used for fitting only one set of data. Furthermore, the non-zero signal in the $Y$ channel, and, in particular, its increase with respect to the signal in the $X$ channel, is related to a retardation of the spin modulation with respect to the rf modulation, which is the essence of the spin inertia effect. This retardation can be described by the phase $\varphi$, so that $\tan\varphi = -Y/X = 2\pi f_\text{m} T_1$, and this relation in principle can be used on its own to determine $T_1$. Note, the shapes of the curves described by Eqs.~\eqref{eq:XSmallW}-\eqref{eq:RSmallW} are determined by $T_1$ only.

The model presented in this section is far from being exact. In particular, we neglect the transverse fluctuations of $\boldsymbol{\omega}_\text{L}$ in the $xy$ plane. Nevertheless, this model illustrates the difference between frozen and time-varying fluctuations of $\boldsymbol{\omega}_\text{L}$. It also provides the analytical expressions for fitting the experimental data which are consistent with more rigorous numerical calculations based on the Bloch equation.  

\section{Discussion}
\begin{figure}
\includegraphics[width=0.8\columnwidth]{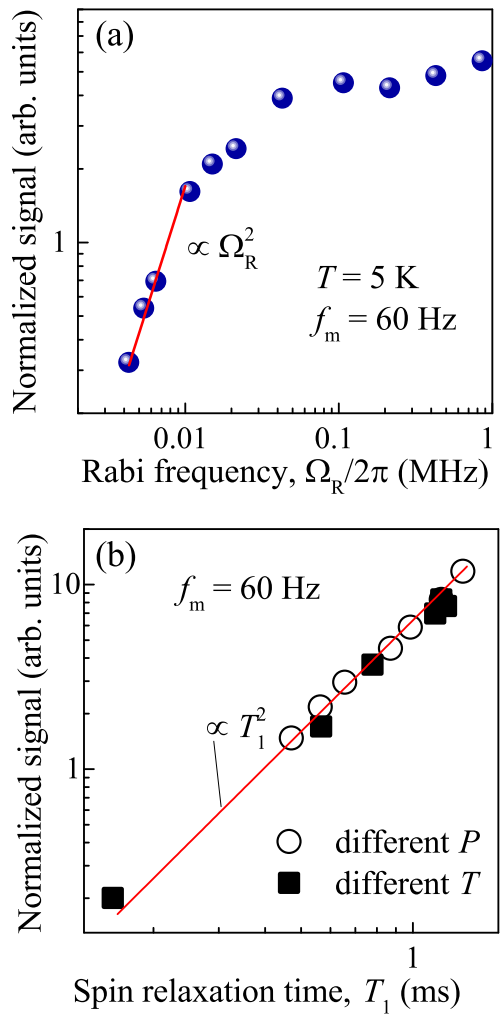}
\caption{(a) Faraday rotation signal normalized to $P$ as a function of the Rabi frequency. $P = 1$~mW, $T = 5$~K. (b) Faraday rotation signal normalized to $P$ as a function of the longitudinal spin relaxation time $T_1$. Full squares correspond to different temperatures. Open circles correspond to different laser powers at $T = 5$~K. (a),(b) Red lines show the quadratic dependence.  $f_\text{m} = 60$~Hz, $B = 4$~mT, $\theta = 12^\circ$, $f_\text{rf} = 80$~MHz.}
\label{fig:DquadDep}
\end{figure}

Thus, we have shown that an rf field acts differently on the spin ensemble depending on whether the Larmor frequencies of individual electrons are constant or varying on a timescale shorter than $T_1$. In the first case, the rf field causes rapid polarization decay (with the rate $\sim \Omega_\text{R}$) for the subensemble of spins whose Larmor frequencies are close enough to the frequency of the rf field. In the second case, the effect of the rf field is more smooth, as it brings an additional contribution to the spin relaxation rate of all spins. The validity of either approach for a certain physical system can be formulated as follows. Time-varying fluctuations approach is valid if the fluctuations correlation time is much shorter than $T_1$, $\tau_\text{c} \ll T_1$, and the spread of these varying fluctuations is much larger than the Rabi frequency, $\Delta\omega_\text{max} \gg \Omega_\text{R}$. Note that $\Delta\omega_\text{max}$ may be smaller than the total width of the spin resonance which may be contributed by both static and time-varying spread of Larmor frequencies, originated, e.g., from $g$ factor spread and nuclear spin fluctuations, respectively. If at least one of the above criteria is not fulfilled, the frozen fluctuations approach should be used.
  
Spin systems with frozen and time-varying fluctuations of the Larmor frequency also manifest themselves differently in the experiment. Apart from quantitative difference reflected in the degree of agreement of the corresponding theoretical fits and the experimental data [Fig.~\ref{fig:EnsTime}(d)], there are also qualitative distinctions between the two. The first is that for small $\Omega_\text{R}$, the signal is linear in $\Omega_\text{R}$ in the case of frozen fluctuations [Eqs.~\eqref{eq:XfrozSmall},\eqref{eq:YfrozSmall}] and is quadratic in $\Omega_\text{R}$ in the case of time-varying fluctuations [Eq.~\eqref{eq:RSmallW}]. The experimental dependence of the signal on $\Omega_\text{R}$, parametrized by the voltage applied to the rf coil, is in fact quadratic for small $\Omega_\text{R}$ [Fig.~\ref{fig:DquadDep}(a)]. The second qualitative distinction is that for low modulation frequencies the signal is linear in $T_1$ in the case of frozen fluctuations [Eqs.~\eqref{eq:XfrozSmall},\eqref{eq:YfrozSmall}] and is quadratic in $T_1$ in the case of time-varying fluctuations [Eq.~\eqref{eq:RSmallW}]. In Fig.~\ref{fig:DquadDep}(b), we plot the normalized signal (measured for the low modulation frequency $f_\text{m} = 60$~Hz) as a function of $T_1$, where different full and open symbols correspond to the different laser powers, and different temperatures, respectively. Note that an increase in both power and temperature leads to a decrease in $T_1$ (Figs.~\ref{fig:PDep} and \ref{fig:TDep}). Figure~\ref{fig:DquadDep} shows the Faraday rotation signal normalized to the laser power $P$, since spin polarization, i.e. Faraday rotation, is obviously proportional to $P$ [Eqs.~\eqref{eq:XfrozSmall},\eqref{eq:YfrozSmall},\eqref{eq:RSmallW}]. We remind that the signal registered from the balanced photodetector through the lock-in amplifier is proportional to the Faraday rotation angle times $P$, since the same laser beam is used also for detection. Thus, Fig.~\ref{fig:DquadDep} shows the pristine signal normalized to $P^2$. The dependence in Fig.~\ref{fig:DquadDep}(b) is again quadratic, which confirms the model of time-varying fluctuations. Note that the existence of fluctuations of internal (nuclear) magnetic fields on a timescale shorter than $T_1$ also follows from the polarization recovery curve for Ce$^{3+}$:YAG \cite{Azamat2017}, where spin polarization at zero external field vanish almost completely. Meanwhile, in the case of for the frozen nuclear fluctuations, the spin polarization at $B = 0$ should drop to 1/3 of its value in high magnetic fields \cite{Merkulov2002}.
Therefore, in addition to providing a method to determine $T_1$ for a given $g$ factor, the technique of resonant spin inertia, also makes it possible to reveal the time-varying nature of the Larmor frequency spread and to estimate the corresponding correlation time of the fluctuations $\tau_\text{c} \sim 0.1$~ms.

\section{Conclusions}
In conclusion, we have developed a $g$-factor-selective method of determining the longitudinal spin relaxation time $T_1$ based on the spin inertia principle. In this method the spin polarization is constantly pumped by a laser. Simultaneously, an rf magnetic field resonant with the Larmor frequency of the studied spin resonance is applied and depolarizes the spin ensemble. The modulation of the rf field results in the corresponding modulation of the spin polarization. By increasing the modulation frequency, one can observe the decrease in the modulation depth of the spin polarization and determine the longitudinal spin relaxation time $T_1$. We have applied this method to the cerium ions embedded at six magnetically inequivalent sites of the YAG lattice which have a strongly anisotropic $g$ tensor for the optically active electron in the 4f level. This results in six magnetic resonances that need to be addressed individually, which is impossible with all-optical methods. We measure millisecond $T_1$, its magnetic field and temperature dependences, and reveal its anisotropy. Moreover, the resonant spin inertia method allows to unveil the nature of the Larmor frequency spread.  
In particular, for the studied Ce$^{3+}$:YAG system, this spread results from the variation of the internal (nuclear) fields on a timescale shorter than 0.1~ms.  

\section{Acknowledgements}
We are grateful to A.~R.~Korotneva and M.~V.~Kravtsov for help with experiments, to D.~V.~Azamat for useful advises, to M.~L.~Skorikov and D.~R.~Yakovlev for fruitful discussions and constructive remarks on the manuscript, and to D. H. Feng for providing the sample. The work was supported by the Ministry of Science and Higher Education of the Russian Federation, Contract No. 075-15-2021-598 at the P. N. Lebedev Physical Institute (experimental studies), and by the Russian Science Foundation through Grant No. 18-72-10073 (development of the technique and modelling).

\begin{widetext}
\section{Appendix A: angular dependence of spin resonances}
\setcounter{equation}{0}
\renewcommand{\theequation}{A\arabic{equation}}
\renewcommand{\thefigure}{A\arabic{figure}}
\renewcommand{\bibnumfmt}[1]{[A#1]}
\renewcommand{\citenumfont}[1]{A#1}
\renewcommand{\thetable}{A\arabic{table}}
\renewcommand{\thesection}{A\arabic{section}}

There are six magnetically inequivalent $c$ sites in the YAG unit cell with D2 point symmetry in which Ce$^{3+}$ ions can be embedded. Directions of $g$-tensor principal axes for these sites are given in Tab.~\ref{tab:axes}. Therefore, for a fixed direction of the magnetic field $\mathbf{B}$, six Larmor frequencies can be calculated:
\begin{equation}
\boldsymbol{\omega}_{\text{L},i} = \frac{\mu_\text{B}}{\hbar}\hat{A}_i \hat{g}_\text{d} \hat{A}_i^{-1} \boldsymbol{B},
\label{eq:ALrm}
\end{equation}
where $\hat{A}_i$ is a transformation matrix from the systems with axes given in Tab.~\ref{tab:axes}, in which $g$ tensor has the diagonal form $\hat{g}_\text{d} = diag(g_{x}, g_{y}, g_{z})$, $g_{x} = 2.738$, $g_{y} = 1.872$, $g_{z} = 0.91$, to the lattice coordinate system ([1,0,0], [0,1,0], [0,0,1]):
\begin{eqnarray}
\hat{A}_1 = \left(\begin{matrix}
0 & 1/\sqrt{2} & 1/\sqrt{2}\\
0 & -1/\sqrt{2} & 1/\sqrt{2}\\
1 & 0 & 0
\end{matrix}\right),
\hat{A}_2 = \left(\begin{matrix}
0 & 1/\sqrt{2} & 1/\sqrt{2}\\
0 & 1/\sqrt{2} & -1/\sqrt{2}\\
-1 & 0 & 0
\end{matrix}\right),
\hat{A}_3 = \left(\begin{matrix}
0 & -1/\sqrt{2} & 1/\sqrt{2}\\
1 & 0 & 0\\
0 & 1/\sqrt{2} & 1/\sqrt{2}
\end{matrix}\right),\nonumber\\
\hat{A}_4 = \left(\begin{matrix}
0 & 1/\sqrt{2} & -1/\sqrt{2}\\
-1 & 0 & 0\\
0 & 1/\sqrt{2} & 1/\sqrt{2}
\end{matrix}\right),
\hat{A}_5 = \left(\begin{matrix}
1 & 0 & 0\\
0 & 1/\sqrt{2} & 1/\sqrt{2}\\
0 & -1/\sqrt{2} & 1/\sqrt{2}
\end{matrix}\right),
\hat{A}_6 = \left(\begin{matrix}
-1 & 0 & 0\\
0 & 1/\sqrt{2} & 1/\sqrt{2}\\
0 & 1/\sqrt{2} & -1/\sqrt{2}
\end{matrix}\right).
\end{eqnarray}

The angle $\theta$ in Figs.~\ref{fig:BDep}(e) and \ref{fig:T1Anisotr}(b) was varied in the plane containing vectors $\begin{pmatrix}0\\0\\1\end{pmatrix}$ (sample normal) and $\begin{pmatrix} \cos\phi_0\\ -\sin\phi_0\\ 0\end{pmatrix}$ (which lies in the sample plane), where $\phi_0 = 24^\circ$ and the vectors are given in the lattice coordinate system. Thus, six spin resonance frequencies are given by the absolute value of $\boldsymbol{\omega}_\text{L}$ from Eq.~\eqref{eq:ALrm}, where in the lattice coordinate system $\mathbf{B} = B\begin{pmatrix}
\cos\phi_0 \sin\theta \\
-\sin\phi_0 \sin\theta \\
\cos\theta
\end{pmatrix}$.

\begin{table}
  \centering
  \begin{tabular}{|p{2cm}|p{2cm}|p{2cm}|p{2cm}|}
    \hline
    Site & x & y & z \\ \hline
    1 & $[0,0,1]$ & $[1,\bar{1},0]$ & $[1,1,0]$ \\ \hline
    2 & $[0,0,\bar{1}]$ & $[1,1,0]$ & $[1,\bar{1},0]$ \\ \hline
    3 & $[0,1,0]$ & $[\bar{1},0,1]$ & $[1,0,1]$ \\ \hline
    4 & $[0,\bar{1},0]$ & $[1,0,1]$ & $[\bar{1},0,1]$ \\ \hline
    5 & $[1,0,0]$ & $[0,1,\bar{1}]$ & $[0,1,1]$ \\ \hline
    6 & $[\bar{1},0,0]$ & $[0,1,1]$ & $[0,1,\bar{1}]$ \\
    \hline
  \end{tabular}
  \caption{Directions of the $g$ tensor principal axes for YAG $c$ sites.}\label{tab:axes}
\end{table}

\section{Appendix B: effect of the rf modulation for frozen field fluctuations}
\setcounter{equation}{0}
\renewcommand{\theequation}{B\arabic{equation}}
\renewcommand{\thefigure}{B\arabic{figure}}
\renewcommand{\bibnumfmt}[1]{[B#1]}
\renewcommand{\citenumfont}[1]{B#1}
\renewcommand{\thetable}{B\arabic{table}}
\renewcommand{\thesection}{B\arabic{section}}
Under modulation of the rf field with a period $T_\text{m}$, the ensemble-averaged  spin dynamics for frozen fluctuations of $\omega_\text{L}$ having Lorentzian distribution $\Delta\omega_\text{max}/\pi(\omega_\text{max}^2+\Delta\omega_\text{L}^2)$ and $T_1=T_2$ is given by the following equations:
\begin{equation}
<S_z>_\text{ens}(t) = P T_1 \\
\times\begin{cases}
1-\beta \exp(-t/T_1),   0<t<T_\text{m}/2\\
1-\alpha - (\beta-\alpha)\exp[(f_\text{m}T_1)^{-1}-t/T_1] + f(t), T_\text{m}/2<t<T_\text{m},
\end{cases}
\label{eq:SxRFEns}
\end{equation}
where $f(t)$ decays with the rate $\sim \Omega_\text{R} \gg 1/T_1$ and unimportant in our case and
\begin{eqnarray}
\alpha = \frac{\Omega_\text{R}}{\Omega_\text{R} + \Delta\omega_\text{max}},\\
\beta = \frac{\Omega_\text{R}}{\Omega_\text{R} + \Delta\omega_\text{max}\sqrt{1-\exp[-(f_\text{m}T_1)^{-1}]}}.
\end{eqnarray}
The convolution of this spin temporal profile with $\sin(2\pi f_\text{m} t)$ and $\cos(2\pi f_\text{m} t)$ gives the signals measured in the $X$ and $Y$ channels of the lock-in amplifier, respectively:
\begin{eqnarray}
X = \frac{P T_1}{\pi} \alpha
-\frac{2\pi P T_1^3 f_\text{m}^2}{1+4\pi^2 T_1^2 f_\text{m}^2}\left\{1+\exp[(2T_1 f_\text{m})^{-1}]\right\}\left\{\alpha - \beta (1 - \exp[-(2T_1 f_\text{m})^{-1}])\right\},
\label{eq:Xfroz}\\
Y = -\frac{P T_1^2 f_\text{m}}{1+4\pi^2 T_1^2 f_\text{m}^2}\left\{1+\exp[(2T_1 f_\text{m})^{-1}]\right\}\left\{\alpha - \beta (1 - \exp[-(2T_1 f_\text{m})^{-1}])\right\}.
\label{eq:Yfroz}
\end{eqnarray}

\section{Appendix C: effect of the rf modulation for time-varying field fluctuations}
\setcounter{equation}{0}
\renewcommand{\theequation}{C\arabic{equation}}
\renewcommand{\thefigure}{C\arabic{figure}}
\renewcommand{\bibnumfmt}[1]{[C#1]}
\renewcommand{\citenumfont}[1]{C#1}
\renewcommand{\thetable}{C\arabic{table}}
\renewcommand{\thesection}{C\arabic{section}}
Under modulation of the rf field with period $T_\text{m}$, the spin dynamics averaged over temporal fluctuations of $\omega_\text{L}$ can be given by the following equations:
\begin{multline}
<S_z>_\text{time}(t) = P T_1-\frac{P T_1}{1+\tau_\text{rf}/T_1}\frac{1-\exp[-(2 f_\text{m} T_1)^{-1}-(2f_\text{m}\tau_\text{rf})^{-1}]}{1-\exp[-(f_\text{m} T_1)^{-1}-(2f_\text{m}\tau_\text{rf})^{-1}]}\exp(-t/T_1),   0<t<T_\text{m}/2;\\
<S_z>_\text{time}(t) = \frac{P}{\tau_\text{rf}^{-1}+T_1^{-1}}\\
+\frac{P T_1}{1+\tau_\text{rf}/T_1}\frac{\exp[(2 f_\text{m} T_1)^{-1}+(2f_\text{m}\tau_\text{rf})^{-1}]-\exp[(2f_\text{m}\tau_\text{rf})^{-1}]}{1-\exp[-(f_\text{m} T_1)^{-1}-(2f_\text{m}\tau_\text{rf})^{-1}]}\exp(-t/T_1-t/\tau_\text{rf}),   T_\text{m}/2<t<T_\text{m}.
\label{eq:SxRFTime}
\end{multline}
The convolution of Eqs.~\eqref{eq:SxRFTime} with $\sin(2\pi f_\text{m} t)$ and $\cos(2\pi f_\text{m} t)$ are given by the following equations, respectively:
\begin{multline}
X = \frac{1}{\pi}\frac{P T_1^2}{T_1+\tau_\text{rf}} \Big\{1-
2\pi^2 f_\text{m}^2
\left[\frac{1}{T_1^{-2}+4\pi^2 f_\text{m}^2}+\frac{1}{(T_1^{-1}+\tau_\text{rf}^{-1})^2+4\pi^2 f_\text{m}^2}\right]\\
-2\pi^2 f_\text{m}^2\frac{1-\exp[-(2f_\text{m}\tau_\text{rf})^{-1}]}{\exp[(2f_\text{m}T_1)^{-1}]-\exp[-(2 f_\text{m} T_1)^{-1}-(2f_\text{m}\tau_\text{rf})^{-1}]}
\left[\frac{1}{T_1^{-2}+4\pi^2 f_\text{m}^2}-\frac{1}{(T_1^{-1}+\tau_\text{rf}^{-1})^2+4\pi^2 f_\text{m}^2}\right]
\Big\};
\label{eq:X}
\end{multline} 
\begin{multline}
Y = -\frac{P T_1^2 f_\text{m}}{T_1+\tau_\text{rf}} \Big\{
\frac{T_1^{-1}}{T_1^{-2}+4\pi^2 f_\text{m}^2}+\frac{T_1^{-1}+\tau_\text{rf}^{-1}}{(T_1^{-1}+\tau_\text{rf}^{-1})^2+4\pi^2 f_\text{m}^2}\\
+\frac{1-\exp[-(2f_\text{m}\tau_\text{rf})^{-1}]}{\exp[(2f_\text{m}T_1)^{-1}]-\exp[-(2 f_\text{m} T_1)^{-1}-(2f_\text{m}\tau_\text{rf})^{-1}]}
\left[\frac{T_1^{-1}}{T_1^{-2}+4\pi^2 f_\text{m}^2}-\frac{T_1^{-1}+\tau_\text{rf}^{-1}}{(T_1^{-1}+\tau_\text{rf}^{-1})^2+4\pi^2 f_\text{m}^2}\right]
\Big\}.
\label{eq:Y}
\end{multline}

\end{widetext}


\begin{thebibliography}{30}%
\makeatletter
\providecommand \@ifxundefined [1]{%
 \@ifx{#1\undefined}
}%
\providecommand \@ifnum [1]{%
 \ifnum #1\expandafter \@firstoftwo
 \else \expandafter \@secondoftwo
 \fi
}%
\providecommand \@ifx [1]{%
 \ifx #1\expandafter \@firstoftwo
 \else \expandafter \@secondoftwo
 \fi
}%
\providecommand \natexlab [1]{#1}%
\providecommand \enquote  [1]{``#1''}%
\providecommand \bibnamefont  [1]{#1}%
\providecommand \bibfnamefont [1]{#1}%
\providecommand \citenamefont [1]{#1}%
\providecommand \href@noop [0]{\@secondoftwo}%
\providecommand \href [0]{\begingroup \@sanitize@url \@href}%
\providecommand \@href[1]{\@@startlink{#1}\@@href}%
\providecommand \@@href[1]{\endgroup#1\@@endlink}%
\providecommand \@sanitize@url [0]{\catcode `\\12\catcode `\$12\catcode
  `\&12\catcode `\#12\catcode `\^12\catcode `\_12\catcode `\%12\relax}%
\providecommand \@@startlink[1]{}%
\providecommand \@@endlink[0]{}%
\providecommand \url  [0]{\begingroup\@sanitize@url \@url }%
\providecommand \@url [1]{\endgroup\@href {#1}{\urlprefix }}%
\providecommand \urlprefix  [0]{URL }%
\providecommand \Eprint [0]{\href }%
\providecommand \doibase [0]{https://doi.org/}%
\providecommand \selectlanguage [0]{\@gobble}%
\providecommand \bibinfo  [0]{\@secondoftwo}%
\providecommand \bibfield  [0]{\@secondoftwo}%
\providecommand \translation [1]{[#1]}%
\providecommand \BibitemOpen [0]{}%
\providecommand \bibitemStop [0]{}%
\providecommand \bibitemNoStop [0]{.\EOS\space}%
\providecommand \EOS [0]{\spacefactor3000\relax}%
\providecommand \BibitemShut  [1]{\csname bibitem#1\endcsname}%
\let\auto@bib@innerbib\@empty
\bibitem [{\citenamefont {Hahn}(1950)}]{Hahn1950}%
  \BibitemOpen
  \bibfield  {author} {\bibinfo {author} {\bibfnamefont {E.~L.}\ \bibnamefont
  {Hahn}},\ }\bibfield  {title} {\bibinfo {title} {{Spin echoes}},\ }\href
  {https://doi.org/10.1103/PhysRev.80.580} {\bibfield  {journal} {\bibinfo
  {journal} {Phys. Rev.}\ }\textbf {\bibinfo {volume} {80}},\ \bibinfo {pages}
  {580} (\bibinfo {year} {1950})}\BibitemShut {NoStop}%
\bibitem [{\citenamefont {Belykh}\ \emph {et~al.}(2021)\citenamefont {Belykh},
  \citenamefont {Korotneva},\ and\ \citenamefont {Yakovlev}}]{BelykhPRL2021}%
  \BibitemOpen
  \bibfield  {author} {\bibinfo {author} {\bibfnamefont {V.~V.}\ \bibnamefont
  {Belykh}}, \bibinfo {author} {\bibfnamefont {A.~R.}\ \bibnamefont
  {Korotneva}},\ and\ \bibinfo {author} {\bibfnamefont {D.~R.}\ \bibnamefont
  {Yakovlev}},\ }\bibfield  {title} {\bibinfo {title} {{Stimulated resonant
  spin amplification reveals millisecond electron spin coherence time of
  rare-earth ions in solids}},\ }\href
  {https://doi.org/10.1103/PhysRevLett.127.157401} {\bibfield  {journal}
  {\bibinfo  {journal} {Phys. Rev. Lett.}\ }\textbf {\bibinfo {volume} {127}},\
  \bibinfo {pages} {157401} (\bibinfo {year} {2021})}\BibitemShut {NoStop}%
\bibitem [{\citenamefont {Greilich}\ \emph {et~al.}(2006)\citenamefont
  {Greilich}, \citenamefont {Yakovlev}, \citenamefont {Shabaev}, \citenamefont
  {Efros}, \citenamefont {Yugova}, \citenamefont {Oulton}, \citenamefont
  {Stavarache}, \citenamefont {Reuter}, \citenamefont {Wieck},\ and\
  \citenamefont {Bayer}}]{Greilich2006}%
  \BibitemOpen
  \bibfield  {author} {\bibinfo {author} {\bibfnamefont {A.}~\bibnamefont
  {Greilich}}, \bibinfo {author} {\bibfnamefont {D.~R.}\ \bibnamefont
  {Yakovlev}}, \bibinfo {author} {\bibfnamefont {A.}~\bibnamefont {Shabaev}},
  \bibinfo {author} {\bibfnamefont {A.~L.}\ \bibnamefont {Efros}}, \bibinfo
  {author} {\bibfnamefont {I.~A.}\ \bibnamefont {Yugova}}, \bibinfo {author}
  {\bibfnamefont {R.}~\bibnamefont {Oulton}}, \bibinfo {author} {\bibfnamefont
  {V.}~\bibnamefont {Stavarache}}, \bibinfo {author} {\bibfnamefont
  {D.}~\bibnamefont {Reuter}}, \bibinfo {author} {\bibfnamefont
  {A.}~\bibnamefont {Wieck}},\ and\ \bibinfo {author} {\bibfnamefont
  {M.}~\bibnamefont {Bayer}},\ }\bibfield  {title} {\bibinfo {title} {{Mode
  locking of electron spin coherences in singly charged quantum dots}},\ }\href
  {https://doi.org/10.1126/science.1128215} {\bibfield  {journal} {\bibinfo
  {journal} {Science (New York, N.Y.)}\ }\textbf {\bibinfo {volume} {313}},\
  \bibinfo {pages} {341} (\bibinfo {year} {2006})}\BibitemShut {NoStop}%
\bibitem [{\citenamefont {Schweiger}\ and\ \citenamefont
  {Jeschke}(2001)}]{Schweiger2001}%
  \BibitemOpen
  \bibfield  {author} {\bibinfo {author} {\bibfnamefont {A.}~\bibnamefont
  {Schweiger}}\ and\ \bibinfo {author} {\bibfnamefont {G.}~\bibnamefont
  {Jeschke}},\ }\href@noop {}  {\bibinfo {title} {{Principles of pulse
  electron paramagnetic resonance}}}\ (\bibinfo  {publisher} {Oxford
  University Press},\ \bibinfo {address} {New York},\ \bibinfo {year}
  {2001})\BibitemShut {NoStop}%
\bibitem [{\citenamefont {Alger}(1968)}]{Alger1968}%
  \BibitemOpen
  \bibfield  {author} {\bibinfo {author} {\bibfnamefont {R.}~\bibnamefont
  {Alger}},\ }\href@noop {} {\bibinfo {title} {{Electron paramagnetic
  resonance: techniques and applications}}}\ (\bibinfo  {publisher}
  {Interscience Publishers},\ \bibinfo {address} {New York},\ \bibinfo {year}
  {1968})\BibitemShut {NoStop}%
\bibitem [{\citenamefont {Poole}(1996)}]{Poole1996}%
  \BibitemOpen
  \bibfield  {author} {\bibinfo {author} {\bibfnamefont {C.}~\bibnamefont
  {Poole}},\ }\href@noop {} {\bibinfo {title} {{Electron spin resonance:
  a comprehensive treatise on experimental techniques}}},\ \bibinfo {edition}
  {2nd}\ ed.\ (\bibinfo  {publisher} {Dover Publications},\ \bibinfo {address}
  {New York},\ \bibinfo {year} {1996})\BibitemShut {NoStop}%
\bibitem [{\citenamefont {Herve}\ and\ \citenamefont
  {Pescia}(1960)}]{Herve1960}%
  \BibitemOpen
  \bibfield  {author} {\bibinfo {author} {\bibfnamefont {J.}~\bibnamefont
  {Herve}}\ and\ \bibinfo {author} {\bibfnamefont {J.}~\bibnamefont {Pescia}},\
  }\bibfield  {title} {\bibinfo {title} {{R{\'{e}}sonance Paramagn{\'{e}}tique
  — Mesure du temps de relaxation T1 par modulation du champ
  radiofr{\'{e}}quence H1 et d{\'{e}}tection de variation d'aimantation selon
  le champ director}},\ }\href@noop {} {\bibfield  {journal} {\bibinfo
  {journal} {Compt. Rend. Acad. Sci.}\ }\textbf {\bibinfo {volume} {251}},\
  \bibinfo {pages} {665} (\bibinfo {year} {1960})}\BibitemShut {NoStop}%
\bibitem [{\citenamefont {Misra}(2006)}]{Misra2006}%
  \BibitemOpen
  \bibfield  {author} {\bibinfo {author} {\bibfnamefont {S.}~\bibnamefont
  {Misra}},\ }\bibfield  {title} {\bibinfo {title} {{Microwave amplitude
  modualtion technique to measure spin-lattice (T1) and spin-spin (T2) relaxation
  times}},\ }in\ \href@noop {} {\bibinfo {booktitle} {Computational and Instrumental Methods in EPR}},\ \bibinfo {editor} {edited by\ \bibinfo {editor}
  {\bibfnamefont {C.~J.}\ \bibnamefont {Bender}}\ and\ \bibinfo {editor}
  {\bibfnamefont {L.~J.}\ \bibnamefont {Berliner}}}\ (\bibinfo  {publisher}
  {Springer-Verlag},\ \bibinfo {address} {New York},\ \bibinfo {year}
  {2006})\BibitemShut {NoStop}%
\bibitem [{\citenamefont {Colton}\ \emph {et~al.}(2004)\citenamefont {Colton},
  \citenamefont {Kennedy}, \citenamefont {Bracker},\ and\ \citenamefont
  {Gammon}}]{Colton2004}%
  \BibitemOpen
  \bibfield  {author} {\bibinfo {author} {\bibfnamefont {J.~S.}\ \bibnamefont
  {Colton}}, \bibinfo {author} {\bibfnamefont {T.~A.}\ \bibnamefont {Kennedy}},
  \bibinfo {author} {\bibfnamefont {A.~S.}\ \bibnamefont {Bracker}},\ and\
  \bibinfo {author} {\bibfnamefont {D.}~\bibnamefont {Gammon}},\ }\bibfield
  {title} {\bibinfo {title} {{Microsecond spin-flip times in n-GaAs measured by
  time-resolved polarization of photoluminescence}},\ }\href
  {https://doi.org/10.1103/PhysRevB.69.121307} {\bibfield  {journal} {\bibinfo
  {journal} {Phys. Rev. B}\ }\textbf {\bibinfo {volume} {69}},\ \bibinfo
  {pages} {121307(R)} (\bibinfo {year} {2004})}\BibitemShut {NoStop}%
\bibitem [{\citenamefont {Akimov}\ \emph {et~al.}(2006)\citenamefont {Akimov},
  \citenamefont {Feng},\ and\ \citenamefont {Henneberger}}]{Akimov2006}%
  \BibitemOpen
  \bibfield  {author} {\bibinfo {author} {\bibfnamefont {I.~A.}\ \bibnamefont
  {Akimov}}, \bibinfo {author} {\bibfnamefont {D.~H.}\ \bibnamefont {Feng}},\
  and\ \bibinfo {author} {\bibfnamefont {F.}~\bibnamefont {Henneberger}},\
  }\bibfield  {title} {\bibinfo {title} {{Electron Spin Dynamics in a
  Self-Assembled Semiconductor Quantum Dot: The Limit of Low Magnetic
  Fields}},\ }\href {https://doi.org/10.1103/PhysRevLett.97.056602} {\bibfield
  {journal} {\bibinfo  {journal} {Phys. Rev. Lett.}\ }\textbf {\bibinfo
  {volume} {97}},\ \bibinfo {pages} {056602} (\bibinfo {year}
  {2006})}\BibitemShut {NoStop}%
\bibitem [{\citenamefont {Fu}\ \emph {et~al.}(2006)\citenamefont {Fu},
  \citenamefont {Yeo}, \citenamefont {Clark}, \citenamefont {Santori},
  \citenamefont {Stanley}, \citenamefont {Holland},\ and\ \citenamefont
  {Yamamoto}}]{Fu2006}%
  \BibitemOpen
  \bibfield  {author} {\bibinfo {author} {\bibfnamefont {K.-M.~C.}\
  \bibnamefont {Fu}}, \bibinfo {author} {\bibfnamefont {W.}~\bibnamefont
  {Yeo}}, \bibinfo {author} {\bibfnamefont {S.}~\bibnamefont {Clark}}, \bibinfo
  {author} {\bibfnamefont {C.}~\bibnamefont {Santori}}, \bibinfo {author}
  {\bibfnamefont {C.}~\bibnamefont {Stanley}}, \bibinfo {author} {\bibfnamefont
  {M.~C.}\ \bibnamefont {Holland}},\ and\ \bibinfo {author} {\bibfnamefont
  {Y.}~\bibnamefont {Yamamoto}},\ }\bibfield  {title} {\bibinfo {title}
  {{Millisecond spin-flip times of donor-bound electrons in GaAs}},\ }\href
  {https://doi.org/10.1103/PhysRevB.74.121304} {\bibfield  {journal} {\bibinfo
  {journal} {Phys. Rev. B}\ }\textbf {\bibinfo {volume} {74}},\ \bibinfo
  {pages} {121304(R)} (\bibinfo {year} {2006})}\BibitemShut {NoStop}%
\bibitem [{\citenamefont {Colton}\ \emph {et~al.}(2007)\citenamefont {Colton},
  \citenamefont {Heeb}, \citenamefont {Schroeder}, \citenamefont {Stokes},
  \citenamefont {Wienkes},\ and\ \citenamefont {Bracker}}]{Colton2007}%
  \BibitemOpen
  \bibfield  {author} {\bibinfo {author} {\bibfnamefont {J.~S.}\ \bibnamefont
  {Colton}}, \bibinfo {author} {\bibfnamefont {M.~E.}\ \bibnamefont {Heeb}},
  \bibinfo {author} {\bibfnamefont {P.}~\bibnamefont {Schroeder}}, \bibinfo
  {author} {\bibfnamefont {A.}~\bibnamefont {Stokes}}, \bibinfo {author}
  {\bibfnamefont {L.~R.}\ \bibnamefont {Wienkes}},\ and\ \bibinfo {author}
  {\bibfnamefont {A.~S.}\ \bibnamefont {Bracker}},\ }\bibfield  {title}
  {\bibinfo {title} {{Anomalous magnetic field dependence of the T1 spin
  lifetime in a lightly doped GaAs sample}},\ }\href
  {https://doi.org/10.1103/PhysRevB.75.205201} {\bibfield  {journal} {\bibinfo
  {journal} {Phys. Rev. B}\ }\textbf {\bibinfo {volume} {75}},\ \bibinfo
  {pages} {205201} (\bibinfo {year} {2007})}\BibitemShut {NoStop}%
\bibitem [{\citenamefont {Linpeng}\ \emph {et~al.}(2016)\citenamefont
  {Linpeng}, \citenamefont {Karin}, \citenamefont {Durnev}, \citenamefont
  {Barbour}, \citenamefont {Glazov}, \citenamefont {Sherman}, \citenamefont
  {Watkins}, \citenamefont {Seto},\ and\ \citenamefont {Fu}}]{Linpeng2016}%
  \BibitemOpen
  \bibfield  {author} {\bibinfo {author} {\bibfnamefont {X.}~\bibnamefont
  {Linpeng}}, \bibinfo {author} {\bibfnamefont {T.}~\bibnamefont {Karin}},
  \bibinfo {author} {\bibfnamefont {M.~V.}\ \bibnamefont {Durnev}}, \bibinfo
  {author} {\bibfnamefont {R.}~\bibnamefont {Barbour}}, \bibinfo {author}
  {\bibfnamefont {M.~M.}\ \bibnamefont {Glazov}}, \bibinfo {author}
  {\bibfnamefont {E.~Y.}\ \bibnamefont {Sherman}}, \bibinfo {author}
  {\bibfnamefont {S.~P.}\ \bibnamefont {Watkins}}, \bibinfo {author}
  {\bibfnamefont {S.}~\bibnamefont {Seto}},\ and\ \bibinfo {author}
  {\bibfnamefont {K.-M.~C.}\ \bibnamefont {Fu}},\ }\bibfield  {title} {\bibinfo
  {title} {{Longitudinal spin relaxation of donor-bound electrons in direct
  band-gap semiconductors}},\ }\href
  {https://doi.org/10.1103/PhysRevB.94.125401} {\bibfield  {journal} {\bibinfo
  {journal} {Phys. Rev. B}\ }\textbf {\bibinfo {volume} {94}},\ \bibinfo
  {pages} {125401} (\bibinfo {year} {2016})}\BibitemShut {NoStop}%
\bibitem [{\citenamefont {Siyushev}\ \emph {et~al.}(2014)\citenamefont
  {Siyushev}, \citenamefont {Xia}, \citenamefont {Reuter}, \citenamefont
  {Jamali}, \citenamefont {Zhao}, \citenamefont {Yang}, \citenamefont {Duan},
  \citenamefont {Kukharchyk}, \citenamefont {Wieck}, \citenamefont {Kolesov},\
  and\ \citenamefont {Wrachtrup}}]{Siyushev2014}%
  \BibitemOpen
  \bibfield  {author} {\bibinfo {author} {\bibfnamefont {P.}~\bibnamefont
  {Siyushev}}, \bibinfo {author} {\bibfnamefont {K.}~\bibnamefont {Xia}},
  \bibinfo {author} {\bibfnamefont {R.}~\bibnamefont {Reuter}}, \bibinfo
  {author} {\bibfnamefont {M.}~\bibnamefont {Jamali}}, \bibinfo {author}
  {\bibfnamefont {N.}~\bibnamefont {Zhao}}, \bibinfo {author} {\bibfnamefont
  {N.}~\bibnamefont {Yang}}, \bibinfo {author} {\bibfnamefont {C.}~\bibnamefont
  {Duan}}, \bibinfo {author} {\bibfnamefont {N.}~\bibnamefont {Kukharchyk}},
  \bibinfo {author} {\bibfnamefont {A.~D.}\ \bibnamefont {Wieck}}, \bibinfo
  {author} {\bibfnamefont {R.}~\bibnamefont {Kolesov}},\ and\ \bibinfo {author}
  {\bibfnamefont {J.}~\bibnamefont {Wrachtrup}},\ }\bibfield  {title} {\bibinfo
  {title} {{Coherent properties of single rare-earth spin qubits}},\ }\href
  {https://doi.org/10.1038/ncomms4895} {\bibfield  {journal} {\bibinfo
  {journal} {Nat. Commun.}\ }\textbf {\bibinfo {volume} {5}},\ \bibinfo {pages}
  {3895} (\bibinfo {year} {2014})}\BibitemShut {NoStop}%
\bibitem [{\citenamefont {Baumberg}\ \emph {et~al.}(1994)\citenamefont
  {Baumberg}, \citenamefont {Awschalom}, \citenamefont {Samarth}, \citenamefont
  {Luo},\ and\ \citenamefont {Furdyna}}]{Baumberg1994}%
  \BibitemOpen
  \bibfield  {author} {\bibinfo {author} {\bibfnamefont {J.~J.}\ \bibnamefont
  {Baumberg}}, \bibinfo {author} {\bibfnamefont {D.~D.}\ \bibnamefont
  {Awschalom}}, \bibinfo {author} {\bibfnamefont {N.}~\bibnamefont {Samarth}},
  \bibinfo {author} {\bibfnamefont {H.}~\bibnamefont {Luo}},\ and\ \bibinfo
  {author} {\bibfnamefont {J.~K.}\ \bibnamefont {Furdyna}},\ }\bibfield
  {title} {\bibinfo {title} {{Spin beats and dynamical magnetization in quantum
  structures}},\ }\href {https://doi.org/10.1103/PhysRevLett.72.717} {\bibfield
   {journal} {\bibinfo  {journal} {Phys. Rev. Lett.}\ }\textbf {\bibinfo
  {volume} {72}},\ \bibinfo {pages} {717} (\bibinfo {year} {1994})}\BibitemShut
  {NoStop}%
\bibitem [{\citenamefont {Zheludev}\ \emph {et~al.}(1994)\citenamefont
  {Zheludev}, \citenamefont {Brummell}, \citenamefont {Harley}, \citenamefont
  {Malinowski}, \citenamefont {Popov}, \citenamefont {Ashenford},\ and\
  \citenamefont {Lunn}}]{Zheludev1994}%
  \BibitemOpen
  \bibfield  {author} {\bibinfo {author} {\bibfnamefont {N.~I.}\ \bibnamefont
  {Zheludev}}, \bibinfo {author} {\bibfnamefont {M.~A.}\ \bibnamefont
  {Brummell}}, \bibinfo {author} {\bibfnamefont {R.~T.}\ \bibnamefont
  {Harley}}, \bibinfo {author} {\bibfnamefont {A.}~\bibnamefont {Malinowski}},
  \bibinfo {author} {\bibfnamefont {S.~V.}\ \bibnamefont {Popov}}, \bibinfo
  {author} {\bibfnamefont {D.~E.}\ \bibnamefont {Ashenford}},\ and\ \bibinfo
  {author} {\bibfnamefont {B.}~\bibnamefont {Lunn}},\ }\bibfield  {title}
  {\bibinfo {title} {{Giant specular inverse Faraday effect in
  Cd$_{0.6}$Mn$_{0.4}$Te}},\ }\href
  {https://doi.org/10.1016/0038-1098(94)90064-7} {\bibfield  {journal}
  {\bibinfo  {journal} {Solid State Commun.}\ }\textbf {\bibinfo {volume}
  {89}},\ \bibinfo {pages} {823} (\bibinfo {year} {1994})}\BibitemShut
  {NoStop}%
\bibitem [{\citenamefont {Belykh}\ \emph {et~al.}(2016)\citenamefont {Belykh},
  \citenamefont {Evers}, \citenamefont {Yakovlev}, \citenamefont {Fobbe},
  \citenamefont {Greilich},\ and\ \citenamefont {Bayer}}]{Belykh2016Ext}%
  \BibitemOpen
  \bibfield  {author} {\bibinfo {author} {\bibfnamefont {V.~V.}\ \bibnamefont
  {Belykh}}, \bibinfo {author} {\bibfnamefont {E.}~\bibnamefont {Evers}},
  \bibinfo {author} {\bibfnamefont {D.~R.}\ \bibnamefont {Yakovlev}}, \bibinfo
  {author} {\bibfnamefont {F.}~\bibnamefont {Fobbe}}, \bibinfo {author}
  {\bibfnamefont {A.}~\bibnamefont {Greilich}},\ and\ \bibinfo {author}
  {\bibfnamefont {M.}~\bibnamefont {Bayer}},\ }\bibfield  {title} {\bibinfo
  {title} {{Extended pump-probe Faraday rotation spectroscopy of the
  submicrosecond electron spin dynamics in n-type GaAs}},\ }\href
  {https://doi.org/10.1103/PhysRevB.94.241202} {\bibfield  {journal} {\bibinfo
  {journal} {Phys. Rev. B}\ }\textbf {\bibinfo {volume} {94}},\ \bibinfo
  {pages} {241202(R)} (\bibinfo {year} {2016})}\BibitemShut {NoStop}%
\bibitem [{\citenamefont {Heisterkamp}\ \emph {et~al.}(2015)\citenamefont
  {Heisterkamp}, \citenamefont {Zhukov}, \citenamefont {Greilich},
  \citenamefont {Yakovlev}, \citenamefont {Korenev}, \citenamefont {Pawlis},\
  and\ \citenamefont {Bayer}}]{Heisterkamp2015}%
  \BibitemOpen
  \bibfield  {author} {\bibinfo {author} {\bibfnamefont {F.}~\bibnamefont
  {Heisterkamp}}, \bibinfo {author} {\bibfnamefont {E.~A.}\ \bibnamefont
  {Zhukov}}, \bibinfo {author} {\bibfnamefont {A.}~\bibnamefont {Greilich}},
  \bibinfo {author} {\bibfnamefont {D.~R.}\ \bibnamefont {Yakovlev}}, \bibinfo
  {author} {\bibfnamefont {V.~L.}\ \bibnamefont {Korenev}}, \bibinfo {author}
  {\bibfnamefont {A.}~\bibnamefont {Pawlis}},\ and\ \bibinfo {author}
  {\bibfnamefont {M.}~\bibnamefont {Bayer}},\ }\bibfield  {title} {\bibinfo
  {title} {{Longitudinal and transverse spin dynamics of donor-bound electrons
  in fluorine-doped ZnSe: Spin inertia versus Hanle effect}},\ }\href
  {https://doi.org/10.1103/PhysRevB.91.235432} {\bibfield  {journal} {\bibinfo
  {journal} {Phys. Rev. B}\ }\textbf {\bibinfo {volume} {91}},\ \bibinfo
  {pages} {235432} (\bibinfo {year} {2015})}\BibitemShut {NoStop}%
\bibitem [{\citenamefont {Zhukov}\ \emph {et~al.}(2018)\citenamefont {Zhukov},
  \citenamefont {Kirstein}, \citenamefont {Smirnov}, \citenamefont {Yakovlev},
  \citenamefont {Glazov}, \citenamefont {Reuter}, \citenamefont {Wieck},
  \citenamefont {Bayer},\ and\ \citenamefont {Greilich}}]{Zhukov2018}%
  \BibitemOpen
  \bibfield  {author} {\bibinfo {author} {\bibfnamefont {E.~A.}\ \bibnamefont
  {Zhukov}}, \bibinfo {author} {\bibfnamefont {E.}~\bibnamefont {Kirstein}},
  \bibinfo {author} {\bibfnamefont {D.~S.}\ \bibnamefont {Smirnov}}, \bibinfo
  {author} {\bibfnamefont {D.~R.}\ \bibnamefont {Yakovlev}}, \bibinfo {author}
  {\bibfnamefont {M.~M.}\ \bibnamefont {Glazov}}, \bibinfo {author}
  {\bibfnamefont {D.}~\bibnamefont {Reuter}}, \bibinfo {author} {\bibfnamefont
  {A.~D.}\ \bibnamefont {Wieck}}, \bibinfo {author} {\bibfnamefont
  {M.}~\bibnamefont {Bayer}},\ and\ \bibinfo {author} {\bibfnamefont
  {A.}~\bibnamefont {Greilich}},\ }\bibfield  {title} {\bibinfo {title} {{Spin
  inertia of resident and photoexcited carriers in singly charged quantum
  dots}},\ }\href {https://doi.org/10.1103/PhysRevB.98.121304} {\bibfield
  {journal} {\bibinfo  {journal} {Phys. Rev. B}\ }\textbf {\bibinfo {volume}
  {98}},\ \bibinfo {pages} {121304(R)} (\bibinfo {year} {2018})}\BibitemShut
  {NoStop}%
\bibitem [{\citenamefont {Smirnov}\ \emph {et~al.}(2018)\citenamefont
  {Smirnov}, \citenamefont {Zhukov}, \citenamefont {Kirstein}, \citenamefont
  {Yakovlev}, \citenamefont {Reuter}, \citenamefont {Wieck}, \citenamefont
  {Bayer}, \citenamefont {Greilich},\ and\ \citenamefont
  {Glazov}}]{Smirnov2018}%
  \BibitemOpen
  \bibfield  {author} {\bibinfo {author} {\bibfnamefont {D.~S.}\ \bibnamefont
  {Smirnov}}, \bibinfo {author} {\bibfnamefont {E.~A.}\ \bibnamefont {Zhukov}},
  \bibinfo {author} {\bibfnamefont {E.}~\bibnamefont {Kirstein}}, \bibinfo
  {author} {\bibfnamefont {D.~R.}\ \bibnamefont {Yakovlev}}, \bibinfo {author}
  {\bibfnamefont {D.}~\bibnamefont {Reuter}}, \bibinfo {author} {\bibfnamefont
  {A.~D.}\ \bibnamefont {Wieck}}, \bibinfo {author} {\bibfnamefont
  {M.}~\bibnamefont {Bayer}}, \bibinfo {author} {\bibfnamefont
  {A.}~\bibnamefont {Greilich}},\ and\ \bibinfo {author} {\bibfnamefont
  {M.~M.}\ \bibnamefont {Glazov}},\ }\bibfield  {title} {\bibinfo {title}
  {{Theory of spin inertia in singly charged quantum dots}},\ }\href
  {https://doi.org/10.1103/PhysRevB.98.125306} {\bibfield  {journal} {\bibinfo
  {journal} {Phys. Rev. B}\ }\textbf {\bibinfo {volume} {98}},\ \bibinfo
  {pages} {125306} (\bibinfo {year} {2018})}\BibitemShut {NoStop}%
\bibitem [{\citenamefont {Belykh}\ \emph
  {et~al.}(2019{\natexlab{a}})\citenamefont {Belykh}, \citenamefont {Yakovlev},
  \citenamefont {Glazov}, \citenamefont {Grigoryev}, \citenamefont {Hussain},
  \citenamefont {Rautert}, \citenamefont {Dirin}, \citenamefont {Kovalenko},\
  and\ \citenamefont {Bayer}}]{Belykh2019}%
  \BibitemOpen
  \bibfield  {author} {\bibinfo {author} {\bibfnamefont {V.~V.}\ \bibnamefont
  {Belykh}}, \bibinfo {author} {\bibfnamefont {D.~R.}\ \bibnamefont
  {Yakovlev}}, \bibinfo {author} {\bibfnamefont {M.~M.}\ \bibnamefont
  {Glazov}}, \bibinfo {author} {\bibfnamefont {P.~S.}\ \bibnamefont
  {Grigoryev}}, \bibinfo {author} {\bibfnamefont {M.}~\bibnamefont {Hussain}},
  \bibinfo {author} {\bibfnamefont {J.}~\bibnamefont {Rautert}}, \bibinfo
  {author} {\bibfnamefont {D.~N.}\ \bibnamefont {Dirin}}, \bibinfo {author}
  {\bibfnamefont {M.~V.}\ \bibnamefont {Kovalenko}},\ and\ \bibinfo {author}
  {\bibfnamefont {M.}~\bibnamefont {Bayer}},\ }\bibfield  {title} {\bibinfo
  {title} {{Coherent spin dynamics of electrons and holes in CsPbBr$_3$
  perovskite crystals}},\ }\href {https://doi.org/10.1038/s41467-019-08625-z}
  {\bibfield  {journal} {\bibinfo  {journal} {Nat. Commun.}\ }\textbf {\bibinfo
  {volume} {10}},\ \bibinfo {pages} {673} (\bibinfo {year}
  {2019}{\natexlab{a}})}\BibitemShut {NoStop}%
\bibitem [{\citenamefont {Kirstein}\ \emph {et~al.}(2021)\citenamefont
  {Kirstein}, \citenamefont {Yakovlev}, \citenamefont {Glazov}, \citenamefont
  {Evers}, \citenamefont {Zhukov}, \citenamefont {Belykh}, \citenamefont
  {Kopteva}, \citenamefont {Kudlacik}, \citenamefont {Nazarenko}, \citenamefont
  {Dirin}, \citenamefont {Kovalenko},\ and\ \citenamefont
  {Bayer}}]{Kirstein2021}%
  \BibitemOpen
  \bibfield  {author} {\bibinfo {author} {\bibfnamefont {E.}~\bibnamefont
  {Kirstein}}, \bibinfo {author} {\bibfnamefont {D.~R.}\ \bibnamefont
  {Yakovlev}}, \bibinfo {author} {\bibfnamefont {M.~M.}\ \bibnamefont
  {Glazov}}, \bibinfo {author} {\bibfnamefont {E.}~\bibnamefont {Evers}},
  \bibinfo {author} {\bibfnamefont {E.~A.}\ \bibnamefont {Zhukov}}, \bibinfo
  {author} {\bibfnamefont {V.~V.}\ \bibnamefont {Belykh}}, \bibinfo {author}
  {\bibfnamefont {N.~E.}\ \bibnamefont {Kopteva}}, \bibinfo {author}
  {\bibfnamefont {D.}~\bibnamefont {Kudlacik}}, \bibinfo {author}
  {\bibfnamefont {O.}~\bibnamefont {Nazarenko}}, \bibinfo {author}
  {\bibfnamefont {D.~N.}\ \bibnamefont {Dirin}}, \bibinfo {author}
  {\bibfnamefont {M.~V.}\ \bibnamefont {Kovalenko}},\ and\ \bibinfo {author}
  {\bibfnamefont {M.}~\bibnamefont {Bayer}},\ }\bibfield  {title} {\bibinfo
  {title} {{Lead‐dominated hyperfine interaction impacting the carrier spin
  dynamics in halide perovskites}},\ }\href
  {https://doi.org/10.1002/adma.202105263} {\bibfield  {journal} {\bibinfo
  {journal} {Adv. Mater.}\ }\textbf {\bibinfo {volume} {34}},\ \bibinfo {pages}
  {2105263} (\bibinfo {year} {2021})}\BibitemShut {NoStop}%
\bibitem [{\citenamefont {Lewis}(1966)}]{Lewis1966}%
  \BibitemOpen
  \bibfield  {author} {\bibinfo {author} {\bibfnamefont {H.~R.}\ \bibnamefont
  {Lewis}},\ }\bibfield  {title} {\bibinfo {title} {{Paramagnetic resonance of
  Ce$^{3+}$ in Yttrium Aluminum Garnet}},\ }\href
  {https://doi.org/10.1063/1.1708247} {\bibfield  {journal} {\bibinfo
  {journal} {J. Appl. Phys.}\ }\textbf {\bibinfo {volume} {37}},\ \bibinfo
  {pages} {739} (\bibinfo {year} {1966})}\BibitemShut {NoStop}%
\bibitem [{\citenamefont {Azamat}\ \emph {et~al.}(2017)\citenamefont {Azamat},
  \citenamefont {Belykh}, \citenamefont {Yakovlev}, \citenamefont {Fobbe},
  \citenamefont {Feng}, \citenamefont {Evers}, \citenamefont {Jastrabik},
  \citenamefont {Dejneka},\ and\ \citenamefont {Bayer}}]{Azamat2017}%
  \BibitemOpen
  \bibfield  {author} {\bibinfo {author} {\bibfnamefont {D.~V.}\ \bibnamefont
  {Azamat}}, \bibinfo {author} {\bibfnamefont {V.~V.}\ \bibnamefont {Belykh}},
  \bibinfo {author} {\bibfnamefont {D.~R.}\ \bibnamefont {Yakovlev}}, \bibinfo
  {author} {\bibfnamefont {F.}~\bibnamefont {Fobbe}}, \bibinfo {author}
  {\bibfnamefont {D.~H.}\ \bibnamefont {Feng}}, \bibinfo {author}
  {\bibfnamefont {E.}~\bibnamefont {Evers}}, \bibinfo {author} {\bibfnamefont
  {L.}~\bibnamefont {Jastrabik}}, \bibinfo {author} {\bibfnamefont
  {A.}~\bibnamefont {Dejneka}},\ and\ \bibinfo {author} {\bibfnamefont
  {M.}~\bibnamefont {Bayer}},\ }\bibfield  {title} {\bibinfo {title} {{Electron
  spin dynamics of Ce$^{3+}$ ions in YAG crystals studied by pulse-EPR and
  pump-probe Faraday rotation}},\ }\href
  {https://doi.org/10.1103/PhysRevB.96.075160} {\bibfield  {journal} {\bibinfo
  {journal} {Phys. Rev. B}\ }\textbf {\bibinfo {volume} {96}},\ \bibinfo
  {pages} {075160} (\bibinfo {year} {2017})}\BibitemShut {NoStop}%
\bibitem [{\citenamefont {Belykh}\ \emph {et~al.}(2020)\citenamefont {Belykh},
  \citenamefont {Sob'yanin},\ and\ \citenamefont {Korotneva}}]{Belykh2020}%
  \BibitemOpen
  \bibfield  {author} {\bibinfo {author} {\bibfnamefont {V.~V.}\ \bibnamefont
  {Belykh}}, \bibinfo {author} {\bibfnamefont {D.~N.}\ \bibnamefont
  {Sob'yanin}},\ and\ \bibinfo {author} {\bibfnamefont {A.~R.}\ \bibnamefont
  {Korotneva}},\ }\bibfield  {title} {\bibinfo {title} {{Resonant spin
  amplification meets electron spin resonance in n-GaAs}},\ }\href
  {https://doi.org/10.1103/PhysRevB.102.075201} {\bibfield  {journal} {\bibinfo
   {journal} {Phys. Rev. B}\ }\textbf {\bibinfo {volume} {102}},\ \bibinfo
  {pages} {075201} (\bibinfo {year} {2020})}\BibitemShut {NoStop}%
\bibitem [{\citenamefont {Bell}\ and\ \citenamefont {Bloom}(1961)}]{Bell1961}%
  \BibitemOpen
  \bibfield  {author} {\bibinfo {author} {\bibfnamefont {W.~E.}\ \bibnamefont
  {Bell}}\ and\ \bibinfo {author} {\bibfnamefont {A.~L.}\ \bibnamefont
  {Bloom}},\ }\bibfield  {title} {\bibinfo {title} {{Optically Driven Spin
  Precession}},\ }\href {https://doi.org/10.1103/PhysRevLett.6.280} {\bibfield
  {journal} {\bibinfo  {journal} {Physical Review Letters}\ }\textbf {\bibinfo
  {volume} {6}},\ \bibinfo {pages} {280} (\bibinfo {year} {1961})}\BibitemShut
  {NoStop}%
\bibitem [{\citenamefont {Saeed}\ \emph {et~al.}(2018)\citenamefont {Saeed},
  \citenamefont {Kuhnert}, \citenamefont {Akimov}, \citenamefont {Korenev},
  \citenamefont {Karczewski}, \citenamefont {Wiater}, \citenamefont
  {Wojtowicz}, \citenamefont {Ali}, \citenamefont {Bhatti}, \citenamefont
  {Yakovlev},\ and\ \citenamefont {Bayer}}]{Saeed2018}%
  \BibitemOpen
  \bibfield  {author} {\bibinfo {author} {\bibfnamefont {F.}~\bibnamefont
  {Saeed}}, \bibinfo {author} {\bibfnamefont {M.}~\bibnamefont {Kuhnert}},
  \bibinfo {author} {\bibfnamefont {I.~A.}\ \bibnamefont {Akimov}}, \bibinfo
  {author} {\bibfnamefont {V.~L.}\ \bibnamefont {Korenev}}, \bibinfo {author}
  {\bibfnamefont {G.}~\bibnamefont {Karczewski}}, \bibinfo {author}
  {\bibfnamefont {M.}~\bibnamefont {Wiater}}, \bibinfo {author} {\bibfnamefont
  {T.}~\bibnamefont {Wojtowicz}}, \bibinfo {author} {\bibfnamefont
  {A.}~\bibnamefont {Ali}}, \bibinfo {author} {\bibfnamefont {A.~S.}\
  \bibnamefont {Bhatti}}, \bibinfo {author} {\bibfnamefont {D.~R.}\
  \bibnamefont {Yakovlev}},\ and\ \bibinfo {author} {\bibfnamefont
  {M.}~\bibnamefont {Bayer}},\ }\bibfield  {title} {\bibinfo {title}
  {{Single-beam optical measurement of spin dynamics in CdTe/(Cd,Mg)Te quantum
  wells}},\ }\href {https://doi.org/10.1103/PhysRevB.98.075308} {\bibfield
  {journal} {\bibinfo  {journal} {Phys. Rev. B}\ }\textbf {\bibinfo
  {volume} {98}},\ \bibinfo {pages} {075308} (\bibinfo {year}
  {2018})}\BibitemShut {NoStop}%
\bibitem [{\citenamefont {Belykh}\ \emph
  {et~al.}(2019{\natexlab{b}})\citenamefont {Belykh}, \citenamefont
  {Yakovlev},\ and\ \citenamefont {Bayer}}]{Belykh2019RPOP}%
  \BibitemOpen
  \bibfield  {author} {\bibinfo {author} {\bibfnamefont {V.~V.}\ \bibnamefont
  {Belykh}}, \bibinfo {author} {\bibfnamefont {D.~R.}\ \bibnamefont
  {Yakovlev}},\ and\ \bibinfo {author} {\bibfnamefont {M.}~\bibnamefont
  {Bayer}},\ }\bibfield  {title} {\bibinfo {title} {{Radiofrequency driving of
  coherent electron spin dynamics in n-GaAs detected by Faraday rotation}},\
  }\href {https://doi.org/10.1103/PhysRevB.99.161205} {\bibfield  {journal}
  {\bibinfo  {journal} {Phys. Rev. B}\ }\textbf {\bibinfo {volume} {99}},\
  \bibinfo {pages} {161205(R)} (\bibinfo {year}
  {2019}{\natexlab{b}})}\BibitemShut {NoStop}%
\bibitem [{\citenamefont {Robbins}(1979)}]{Robbins1979}%
  \BibitemOpen
  \bibfield  {author} {\bibinfo {author} {\bibfnamefont {D.~J.}\ \bibnamefont
  {Robbins}},\ }\bibfield  {title} {\bibinfo {title} {{The Effects of Crystal
  Field and Temperature on the Photoluminescence Excitation Efficiency of Ce3+
  in YAG}},\ }\href {https://doi.org/10.1149/1.2129328} {\bibfield  {journal}
  {\bibinfo  {journal} {J. Electrochem. Soc.}\ }\textbf
  {\bibinfo {volume} {126}},\ \bibinfo {pages} {1550} (\bibinfo {year}
  {1979})}\BibitemShut {NoStop}%
\bibitem [{\citenamefont {Kolesov}\ \emph {et~al.}(2013)\citenamefont
  {Kolesov}, \citenamefont {Xia}, \citenamefont {Reuter}, \citenamefont
  {Jamali}, \citenamefont {St{\"{o}}hr}, \citenamefont {Inal}, \citenamefont
  {Siyushev},\ and\ \citenamefont {Wrachtrup}}]{Kolesov2013}%
  \BibitemOpen
  \bibfield  {author} {\bibinfo {author} {\bibfnamefont {R.}~\bibnamefont
  {Kolesov}}, \bibinfo {author} {\bibfnamefont {K.}~\bibnamefont {Xia}},
  \bibinfo {author} {\bibfnamefont {R.}~\bibnamefont {Reuter}}, \bibinfo
  {author} {\bibfnamefont {M.}~\bibnamefont {Jamali}}, \bibinfo {author}
  {\bibfnamefont {R.}~\bibnamefont {St{\"{o}}hr}}, \bibinfo {author}
  {\bibfnamefont {T.}~\bibnamefont {Inal}}, \bibinfo {author} {\bibfnamefont
  {P.}~\bibnamefont {Siyushev}},\ and\ \bibinfo {author} {\bibfnamefont
  {J.}~\bibnamefont {Wrachtrup}},\ }\bibfield  {title} {\bibinfo {title}
  {{Mapping spin coherence of a single rare-earth ion in a crystal onto a
  single photon polarization state}},\ }\href
  {https://doi.org/10.1103/PhysRevLett.111.120502} {\bibfield  {journal}
  {\bibinfo  {journal} {Phys. Rev. Lett.}\ }\textbf {\bibinfo {volume} {111}},\
  \bibinfo {pages} {120502} (\bibinfo {year} {2013})}\BibitemShut {NoStop}%
\bibitem [{\citenamefont {Zych}\ \emph {et~al.}(2000)\citenamefont {Zych},
  \citenamefont {Brecher},\ and\ \citenamefont {Glodo}}]{Zych2000}%
  \BibitemOpen
  \bibfield  {author} {\bibinfo {author} {\bibfnamefont {E.}~\bibnamefont
  {Zych}}, \bibinfo {author} {\bibfnamefont {C.}~\bibnamefont {Brecher}},\ and\
  \bibinfo {author} {\bibfnamefont {J.}~\bibnamefont {Glodo}},\ }\bibfield
  {title} {\bibinfo {title} {{Kinetics of cerium emission in a YAG:Ce single
  crystal: the role of traps}},\ }\href
  {https://doi.org/10.1088/0953-8984/12/8/336} {\bibfield  {journal} {\bibinfo
  {journal} {J. Phys. Condens. Matter}\ }\textbf {\bibinfo {volume}
  {12}},\ \bibinfo {pages} {1947} (\bibinfo {year} {2000})}\BibitemShut
  {NoStop}%
\bibitem [{\citenamefont {Liang}\ \emph {et~al.}(2017)\citenamefont {Liang},
  \citenamefont {Hu}, \citenamefont {Chen}, \citenamefont {Belykh},
  \citenamefont {Jia}, \citenamefont {Sun}, \citenamefont {Feng}, \citenamefont
  {Yakovlev},\ and\ \citenamefont {Bayer}}]{Liang2017}%
  \BibitemOpen
  \bibfield  {author} {\bibinfo {author} {\bibfnamefont {P.}~\bibnamefont
  {Liang}}, \bibinfo {author} {\bibfnamefont {R.~R.}\ \bibnamefont {Hu}},
  \bibinfo {author} {\bibfnamefont {C.}~\bibnamefont {Chen}}, \bibinfo {author}
  {\bibfnamefont {V.~V.}\ \bibnamefont {Belykh}}, \bibinfo {author}
  {\bibfnamefont {T.~Q.}\ \bibnamefont {Jia}}, \bibinfo {author} {\bibfnamefont
  {Z.~R.}\ \bibnamefont {Sun}}, \bibinfo {author} {\bibfnamefont {D.~H.}\
  \bibnamefont {Feng}}, \bibinfo {author} {\bibfnamefont {D.~R.}\ \bibnamefont
  {Yakovlev}},\ and\ \bibinfo {author} {\bibfnamefont {M.}~\bibnamefont
  {Bayer}},\ }\bibfield  {title} {\bibinfo {title} {{Room-temperature electron
  spin dynamics of Ce$^{3+}$ ions in a YAG crystal}},\ }\href
  {https://doi.org/10.1063/1.4984232} {\bibfield  {journal} {\bibinfo
  {journal} {Appl. Phys. Lett.}\ }\textbf {\bibinfo {volume} {110}},\ \bibinfo
  {pages} {222405} (\bibinfo {year} {2017})}\BibitemShut {NoStop}%
\bibitem [{\citenamefont {Merkulov}\ \emph {et~al.}(2002)\citenamefont
  {Merkulov}, \citenamefont {Efros},\ and\ \citenamefont
  {Rosen}}]{Merkulov2002}%
  \BibitemOpen
  \bibfield  {author} {\bibinfo {author} {\bibfnamefont {I.~A.}\ \bibnamefont
  {Merkulov}}, \bibinfo {author} {\bibfnamefont {A.~L.}\ \bibnamefont
  {Efros}},\ and\ \bibinfo {author} {\bibfnamefont {M.}~\bibnamefont {Rosen}},\
  }\bibfield  {title} {\bibinfo {title} {{Electron spin relaxation by nuclei in
  semiconductor quantum dots}},\ }\href
  {https://doi.org/10.1103/PhysRevB.65.205309} {\bibfield  {journal} {\bibinfo
  {journal} {Phys. Rev. B}\ }\textbf {\bibinfo {volume} {65}},\ \bibinfo
  {pages} {205309} (\bibinfo {year} {2002})}\BibitemShut {NoStop}%
\bibitem [{\citenamefont {Bloch}(1946)}]{Bloch1946}%
  \BibitemOpen
  \bibfield  {author} {\bibinfo {author} {\bibfnamefont {F.}~\bibnamefont
  {Bloch}},\ }\bibfield  {title} {\bibinfo {title} {{Nuclear Induction}},\
  }\href {https://doi.org/10.1103/PhysRev.70.460} {\bibfield  {journal}
  {\bibinfo  {journal} {Phys. Rev.}\ }\textbf {\bibinfo {volume} {70}},\
  \bibinfo {pages} {460} (\bibinfo {year} {1946})}\BibitemShut {NoStop}%
\bibitem [{\citenamefont {Abragam}(1961)}]{Abragam1961}%
  \BibitemOpen
  \bibfield  {author} {\bibinfo {author} {\bibfnamefont {A.}~\bibnamefont
  {Abragam}},\ }\href@noop {} {\bibinfo {title} {{Principles of nuclear
  magnetism}}}\ (\bibinfo  {publisher} {Clarendon Press},\ \bibinfo {year}
  {1961})\BibitemShut {NoStop}%
\end{thebibliography}
\end{document}